\documentclass{aa}  
\usepackage{graphicx}
\usepackage{subcaption}
\usepackage{multicol}
\usepackage{multirow}
\usepackage{natbib}
\usepackage{txfonts}
\usepackage[pdfencoding=auto,psdextra]{hyperref}
\hypersetup{
    colorlinks=true,
    linkcolor=red,
    anchorcolor=red,
    filecolor=magenta,      
    urlcolor=magenta,
    citecolor=blue
}
\begin{document}

   \title{AMICO-WL: an optimal filtering algorithm for galaxy cluster detections with weak lensing}


   \author{L. Trobbiani
          \inst{1}\fnmsep\inst{2}\thanks{leonardo.trobbiani2@unibo.it}
          ,
          M. Maturi\inst{3,4}
          ,
          C. Giocoli\inst{2}\fnmsep\inst{5}
          ,
          L. Moscardini\inst{1}\fnmsep\inst{2}\fnmsep\inst{5}
          \and
          G. Panebianco\inst{1}\fnmsep\inst{2}
          }

   \institute{Dipartimento di Fisica e Astronomia "A. Righi", Alma Mater Studiorum Universit\`a di Bologna, Via Gobetti 93/2, I-40129 Bologna, Italy
         \and
             INAF, Osservatorio di Astrofisica e Scienza dello Spazio di Bologna, Via Gobetti 93/3, I-40129 Bologna, Italy
         \and
 Zentrum f\"ur Astronomie, Universitat\"at Heidelberg, Philosophenweg 12, D-69120 Heidelberg, Germany 
    \and
        Institute for Theoretical Physics, Philosophenweg 16, D-69120 Heidelberg, Germany 
         \and
            INFN, Sezione di Bologna, viale Berti Pichat 6/2, I-40127 Bologna, Italy
             }

   \date{Received ...; accepted ....}

 
  \abstract
   {The detection of galaxy clusters, the most massive bounded structures in the universe, is crucial for cosmological analysis. Weak lensing signals allow us to track the distribution of all (dark and baryonic) matter regardless of its observable electromagnetic properties. Upcoming wide-field surveys like Euclid and LSST-Rubin will provide enhanced shape measurements of billions of background galaxies, presenting an unparalleled opportunity to detect galaxy clusters on a vast cosmic scale.}
{The immense data volume generated by these surveys will require efficient and accurate analysis techniques. In this work, we introduce \texttt{AMICO-WL}, an extension of the optimal filtering algorithm implemented in AMICO, a well-tested code developed for optical cluster detection. \texttt{AMICO-WL} implements a specific linear optimal matched filter for weak lensing data in the \texttt{AMICO} infrastructure, using parallelisation and adding an efficient signal-to-noise ratio thresholding approach to set a desired sample purity and a cleaning procedure to deal with blended detections.}
   {The algorithm has been tested on a 25 deg$^2$ field of Euclid-like mock galaxy catalogue with the simulated shear signal produced using DUSTGRAIN-{\em pathfinder} past-light-cones. We implemented a foreground removal procedure based on different cuts of low redshift galaxies from the input catalogue. To evaluate the performance of the method, we used an efficient matching procedure based on the `blinking' of the simulation's individual redshift lensing planes.}
   {Cross-matching the \texttt{AMICO-WL} detections with the dark matter halo sample in the simulation having $M_{200} > 5 \times 10^{13}$ $M_{\odot}/h$ and considering a purity level of $\sim70\%$, the application of the foreground removal doubles the completeness from 6.5\% to 13\% and at the same time produces a significant decrease of spurious detections. }
   {}

\authorrunning{Trobbiani et al.}
\titlerunning{AMICO-WL, detecting clusters via weak gravitational lensing}

\maketitle
%

\section{Introduction}

Galaxy clusters serve as powerful tools in cosmology, acting as indicators of density peaks in the large-scale matter distribution (see, e.g., \citealp{allen11} for a comprehensive review). Their relevance lies in the potential to constrain cosmological parameters through their number density (see, for instance, \citealp{rosati2002evolution}; \citealp{vikhlinin2009chandra}; \citealp{rozo10}; \citealp{costanzi2019methods}; \citealp{lesci2022amicoa}; \citealp{ghirardini2024srg}; \citealp{seppi2024srg}) and spatial distribution (as explored by \citealp{veropalumbo2014improved}; \citealp{marulli2018xxl}; \citealp{to2021dark}; \citealp{lesci2022amicob}; \citealp{romanello2024amico}). Moreover, galaxy clusters can be used to study astrophysical processes influencing galaxy formation and evolution. The impact of clusters and galactic environments on the development of galaxy properties is evident in observed differences between field galaxies and those located in denser environments (e.g. \citealp{dressler1980galaxy}; \citealp{kuchner2017effects}; \citealp{george2011galaxies}). The combined study of both astrophysical properties of clusters and cosmology is needed, as the two fields are deeply interconnected.

Galaxy clusters are predominantly composed of dark matter (approximately $80\%$), which forms haloes. About $15\%$ of the cluster’s mass is ionised hot gas within the intracluster medium (ICM), while stars and gas within the galaxies contribute to only $5\%$ of the total mass.

Large-scale surveys have identified thousands of galaxy clusters through various methods. X-ray observations have detected the hot ICM emission \citep[e.g.,][]{bohringer2001rosat, pacaud2016xxl, piffaretti2011mcxc, sadibekova2024mcxc}, while the Sunyaev-Zel'dovich (SZE) effect has been observed at millimetre wavelengths \citep{schaffer2011first, hasselfield2013atacama, bleem2015galaxy, aghanim2011planck, aiola2020atacama, bleem2020sptpol, koulouridis2021x}. Optical and near-infrared surveys have also been successful in identifying clusters, either by detecting overdensities of galaxies \citep{kepner1999automated, ramella2001finding, sarron2018evolution, werner2023s} or by exploiting known properties of the clusters, such as the luminosity or colour distribution of their galaxies \citep{koester2007maxbcg, gilbank2011red, rykoff2014redmapper, farrens2011optical, bellagamba2018amico, maturi2019amico, eisenhardt2008clusters, andreon2009jkcs, rettura2014candidate, adam19}.

However, these methods primarily rely on baryonic observables (X-ray luminosity, SZE flux, optical richness), which only account for a small fraction of the total cluster mass. This reliance on baryonic matter can introduce selection biases, such as favouring cool-core clusters with centrally concentrated X-ray emission \citep{planck1_11, andreon2011x, eckert2011cool, andreon2024observed}. Brighter clusters are more easily detected, while fainter ones may be missed. By using complementary methods that incorporate independent observables, such as weak lensing, these biases can be mitigated, enabling the detection of both brighter and fainter clusters with greater reliability.

Clusters induce gravitational lensing effects on the light emitted by background galaxies \citep[see e.g.][]{schneider1992properties, pyne1995null}. In the weak lensing regime, the distortion of background galaxy shapes is detectable only statistically \citep[see e.g.][]{schneider2002analysis, pires2020euclid}. Weak-lensing cluster detection stands out by using matter concentration as a tracer, independent of the physical state of the baryonic component. This approach allows the detection of clusters regardless of their emission (X-rays, optical) or scattering (SZ) properties. However, the sensitivity of cluster detection through weak lensing critically relies on the presence of background sources available for lensing.

In the early stages of optical surveys, one of the first techniques for the detection of galaxy clusters through weak lensing signals was the aperture mass (AM) method, introduced by \citet{schneider1998new}. This approach convolved the lensing signal with a filter function that depends on a specific scale, providing a foundation for subsequent advancements in cluster detection techniques. Around the same time, a straightforward Gaussian filter applied to the reconstructed convergence also gained popularity due to its simplicity and effectiveness in detecting clusters \citep[e.g.][]{white2002completeness, miyazaki2002searching, hamana2004searching, tang2005effects, gavazzi2007weak, miyazaki2007subaru, fan2010noisy, shan2012weak, shan2018kids}. These early methods led to the development of more refined approaches, including optimisations of the AM method through the design of various filters. These filters have since been rigorously tested on simulations and applied to numerous optical surveys \citep[e.g.][]{schneider1996detection, schneider1998new, jarvis2004skewness, schirmer2004gabods, hetterscheidt2005searching, hennawi2005shear, maturi2005optimal, maturi2007searching, pace2007testing, schirmer2007gabods, dietrich2010cosmology, hamana2012scatter, lin2016new, miyazaki2018large, hamana2020weak, oguri2021hundreds}. More advanced filtering techniques included the full information on the expected shape of the halo profile in the kernel of the filter. This improvement aimed to reduce the contamination by shape noise and the lensing contribution of the large-scale structure (LSS) to the error (e.g., \citealp{hennawi2005shear}; \citealp{maturi2005optimal}; \citealp{wittman2006first}). 

In this study, we present and test the performances of \texttt{AMICO-WL}, an augmentation of the Adaptive Matched Identifier of Clustered Objects (\texttt{AMICO}) (\citealp{bellagamba2018amico}; \citealp{maturi2019amico}) that includes a weak-gravitational lensing branch for the detection of galaxy clusters. 
\texttt{AMICO} is a representative of the class of linear optimal matched filters; this algorithm is effective in extracting signals with a maximised signal-to-noise ratio ($SNR$). \texttt{AMICO} was officially adopted as one of two algorithms for optical cluster detection by the ESA Euclid mission\footnote{\url{http://sci.esa.int/euclid/}} (\citealp{laureijs2011euclid}; \citealp{mellier2024euclid}).
Its performance stood out in the Euclid Cluster Finder Challenge (\citealp{adam19}), demonstrating good completeness and purity as a function of redshift when applied to mock galaxy photometric catalogues replicating the expected properties of Euclid photometric catalogues. The \texttt{AMICO} algorithm has already demonstrated success in wide-field and/or deep surveys such as the Kilo-Degree Survey (KiDS\footnote{\url{http://kids.strw.leidenuniv.nl/}}; \citealp{de2017third}), J-PAS (\citealp{maturi2023minijpas}) and COSMOS (\citealp{toni2024amico}). The analysis of the KiDS data resulted in a cluster sample (\citealp{maturi2019amico}) that has been used in various cosmological studies \citep[e.g.][]{giocoli2021amico, ingoglia2022amico, lesci2022amicoa, lesci2022amicob} and investigations of galaxy properties in cluster environments \citep[e.g.][]{radovich2020amico, puddu2021amico, castignani2022star, castignani2023star}.

\texttt{AMICO-WL} implements an adjusted version of the optimal linear matched filter defined in
 \cite{maturi2005optimal}. The filter assumes a mean radial profile of the halo shear pattern and a power spectrum for the noise. The efficient \texttt{AMICO} infrastructure is adapted for weak lensing data, for example, by changing the cleaning procedure. Moreover, the pipeline is complemented with new tools developed to increase the reliability of the detection, such as a thresholding method based on the $SNR$ analysis of the map peaks and an effective foreground removal procedure for the reduction of noise from galaxies between the lens and the observer. 

In this paper, we use as a case study the application of \texttt{AMICO-WL} to mock data from dedicated light-cone simulations, replicating the expected source density of the Euclid wide-field survey (\citealp{scaramella22}). The Euclid mission, with its wide-field survey covering approximately 14,000 $\mathrm{deg^{2}}$, will measure the shapes of billions of background galaxies, making the weak-lensing probe sensitive to the detection of clusters. For this reason, the Euclid Collaboration initiated the Weak Lensing Selected Cluster Challenge, a platform to compare different analysis methods. \texttt{AMICO-WL} stands as a key contender among the participating algorithms. In the second phase of the challenge, using twelve 100 deg$^2$ fields of simulated Euclid data, \texttt{AMICO-WL} demonstrated good performance, achieving significant results in purity and completeness. 
The comprehensive results of this challenge will be presented in Manjón-García et al. (in preparation). 

In this work, we quantify the performances of \texttt{AMICO-WL} by implementing a new matching procedure based on the `blinking' of redshift planes in the simulated data to investigate the nature of the detections and the functioning of the detection method in more detail.

This paper is organised as follows. In 
Sect.~\ref{Lensing}, we summarise the fundamental points of gravitational lensing theory. The simulated dataset and reference catalogue are presented in Sect.~\ref{WLsim}. We describe the optimal filtering technique on which \texttt{AMICO-WL} is based and the foreground removal cleaning applied to the dataset in Sect.~\ref{OptFilt}. Section~\ref{AMICOWLapp} shows the steps of the \texttt{AMICO-WL} algorithm, from the creation of the map to the computation of the $SNR$ threshold and the cleaning procedure. The results of the application of \texttt{AMICO-WL} to the mock data are evaluated in Sect.~\ref{StaAna}. In Sect.~\ref{spurious}, we discuss the nature of the spurious detections. Finally, in Sect.~\ref{conclusions}, we outline our conclusions. 
   
\section{Weak Gravitational Lensing}\label{Lensing}
\subsection{Theory}



In this section, we briefly summarise the aspects of weak gravitational lensing relevant to the present study. For a more detailed description of the lensing theory, we refer to \citet{bartelmann2001weak}.


Let us consider a lens at the angular position vector $\boldsymbol{\theta}$ on the sky. We can write the surface mass density as:
\begin{equation}
    \Sigma(\boldsymbol{\theta})=\int_{-\infty}^{+\infty}\rho(\boldsymbol{\theta,z})dz\,,
\end{equation}
where $\rho$ is the mass density, from which we can define the lensing convergence as:
\begin{equation}
\kappa(\boldsymbol{\theta})\equiv\frac{\Sigma(\boldsymbol{\theta})}{\Sigma_{\rm crit}}\,,
\end{equation}
with the critical surface density $\Sigma_{\rm crit}$ defined as 
\begin{equation}
    \Sigma_{\rm crit}\equiv\frac{c^2}{4\pi G}\frac{D_{ls}}{D_l D_s}\,.
\end{equation}
The angular diameter distances $D_s$, $D_l$, and $D_{ls}$ refer to the distances between the observer and the source, 
the observer and the lens, and between the lens and the source, respectively. 

Using the two-dimensional Poisson equation, the relation between the lensing potential $\psi$, a scalar field encoding the light deflection, and the convergence is given by
\begin{equation}
\nabla^2_{\boldsymbol{\theta}}\psi(\boldsymbol{\theta})=2\kappa\,.
\end{equation}
The deflection angle,
\begin{equation}
    \boldsymbol{\alpha}(\boldsymbol{\theta})\equiv \nabla_{\boldsymbol{\theta}}\psi(\boldsymbol{\theta})\,,
\end{equation}
expresses the difference between the true angular position in the source plane $\boldsymbol{\beta}$, and the observed position $\boldsymbol{\theta}$, through the lens equation:
\begin{equation}
    \boldsymbol{\beta}=\boldsymbol{\theta} - \boldsymbol{\alpha}(\boldsymbol{\theta})\,.
\end{equation}
Because of this angular distortion, the shapes of the lensed images differ from the intrinsic shape of the source because light bundles are deflected differentially.
If a source is much smaller than the angular scale on which the lensing properties change, the lens mapping can be linearised. The lens equation in the first-order lens mapping can be written as:
\begin{equation}
    \boldsymbol{\beta}=\mathcal{A}(\boldsymbol{\theta})\boldsymbol{\theta}\,,
\end{equation}
where the Jacobian matrix $\mathcal{A}$ describes the distortion of the images:
\begin{equation}
    \mathcal{A}(\boldsymbol{\theta}) = \frac{\partial\boldsymbol{\beta}}{\partial\boldsymbol{\theta}}= \left( \delta_{ij}- \frac{\partial^2\psi(\boldsymbol{\theta})}{\partial\boldsymbol{\theta}_i\partial\boldsymbol{\theta}_j}\right) = \begin{pmatrix}
    1 -\kappa-\gamma_1 & -\gamma_2 \\
    -\gamma_2          & 1-\kappa+\gamma_1
    \end{pmatrix}\,.
\end{equation}
In the previous equation, the quantities
\begin{equation}\label{gamma1}
    \gamma_1\equiv\frac{1}{2}(\partial_1^2-\partial_2^2)\Psi 
\end{equation}
and
\begin{equation}\label{gamma2}
    \gamma_2\equiv\partial_1\partial_2\Psi 
\end{equation}
represent the two components of the pseudo-vector shear $\gamma$, and the convergence $\kappa$ can be re-written as
\begin{equation}\label{eqn:kappa}
    \kappa\equiv \frac{1}{2}\left(\partial_1^2+\partial_2^2\right)\Psi \ .
\end{equation}
From the previous equations, where we used the Einstein derivative notation, it is clear that $\kappa$ describes the isotropic contraction or dilatation of an image while $\gamma$ is the anisotropic distortion.

Considering two independent modes, $E$ and $B$, to which we can associate two scalar fields $\Psi_E$ and $\Psi_B$, 
we can rewrite the shear components as follows:
\begin{equation}
    \label{eqn:gamma1_EB}
    \gamma_1 = \frac{1}{2}(\Psi_{E,11}-\Psi_{E,22}) -\Psi_{B,12}\,,
\end{equation}
and
\begin{equation}
    \label{eqn:gamma2_EB}
    \gamma_2 = \frac{1}{2}(\Psi_{B,11}-\Psi_{B,22})+\Psi_{E,12}\,.
\end{equation}
Comparing the two expressions above  with 
Eqs. (\ref{gamma1}) and (\ref{gamma2}) 
we note that $\Psi_E(\boldsymbol{\theta}) = \Psi(\boldsymbol{\theta})$ and $\Psi_B(\boldsymbol{\theta}) = 0$, i.e. 
for gravitational lensing the $E$-mode field is related to the shear $\gamma$, while the $B$-mode field  is equal to zero.



The spin-2 shear components, $\gamma_1$ and $\gamma_2$ are defined with respect to a reference Cartesian coordinate frame. Therefore, it is useful to consider these components with respect to a given reference point, such as the cluster centre, to define the tangential $\gamma_+(\boldsymbol{\theta'};\boldsymbol{\theta})$ and cross $\gamma_{\times}(\boldsymbol{\theta'};\boldsymbol{\theta})$ components of the shear at a given position $\boldsymbol{\theta}'$ with respect to $\boldsymbol{\theta}$.
These quantities are directly observable in the weak-lensing regime, where $\kappa\ll1$, $\gamma\ll1$. Averaging the measured values within a circle of radius $\vartheta$ around $\boldsymbol{\theta}$, we can write:
\begin{equation}
\label{eqn:gammaTangAve}
    \langle \gamma_+\rangle(\vartheta;\boldsymbol{\theta})=\frac{\Delta\Sigma}{\Sigma_{\rm crit}}(\vartheta;\boldsymbol{\theta})\,,
\end{equation}
and
\begin{equation}
\label{eqn:gammaCrossAve}
    \langle\gamma_{\times}\rangle(\vartheta;\boldsymbol{\theta})=0\,,
\end{equation}
where $\Delta\Sigma(\vartheta;\boldsymbol{\theta})$ represents the excess surface mass density \citep{umetsu2020cluster}.
The previous equation shows that the average cross-component of the shear measures the $B$-mode distortion pattern, and it is expected to be statistically consistent with zero when the signal is due to weak gravitational lensing distortion only. This highlights that the measurement of the $B$-mode signal provides a useful null test against the presence of systematic uncertainties due to correlated noise. 

\subsection{Weak lensing of galaxies}\label{WL_ell}
In the weak lensing regime, a source with intrinsic ellipticity $\epsilon_s$ displays an image with ellipticity 
\begin{equation}
    \label{eqn:ellipticity}
    \epsilon=\frac{\epsilon_s+g}{1+gì^*\epsilon_s}\,,
\end{equation}
where $g$ represents the reduced shear, defined as:
\begin{equation}
    g(\boldsymbol{\theta})\equiv\frac{\gamma(\boldsymbol{\theta})}{1-\kappa(\boldsymbol{\theta})}\,.
\end{equation}
Equation (\ref{eqn:ellipticity}) shows that the reduced shear determines the observed elongation, combining the effect of lensing and the intrinsic shape of the sources. Since background galaxies are expected to be randomly oriented, the average value of intrinsic source ellipticity $\langle \epsilon_s \rangle$ is expected to be close to zero. Thus, the average value of the ellipticity of the images represents an estimate of the reduced shear: $\langle \epsilon \rangle \simeq g$. Furthermore, since we are in the weak-lensing regime and $\kappa\ll1$ thus $g\simeq\gamma$, we can conclude that the ellipticity of a galaxy image is an unbiased estimate of the local shear. We can then define the tangential $\epsilon_+$ and cross $\epsilon_{\times}$ components of the ellipticity in the same reference frame as $\gamma_+$ and cross $\gamma_{\times}$.

\begin{figure*}
\centering
\includegraphics[width=\hsize]{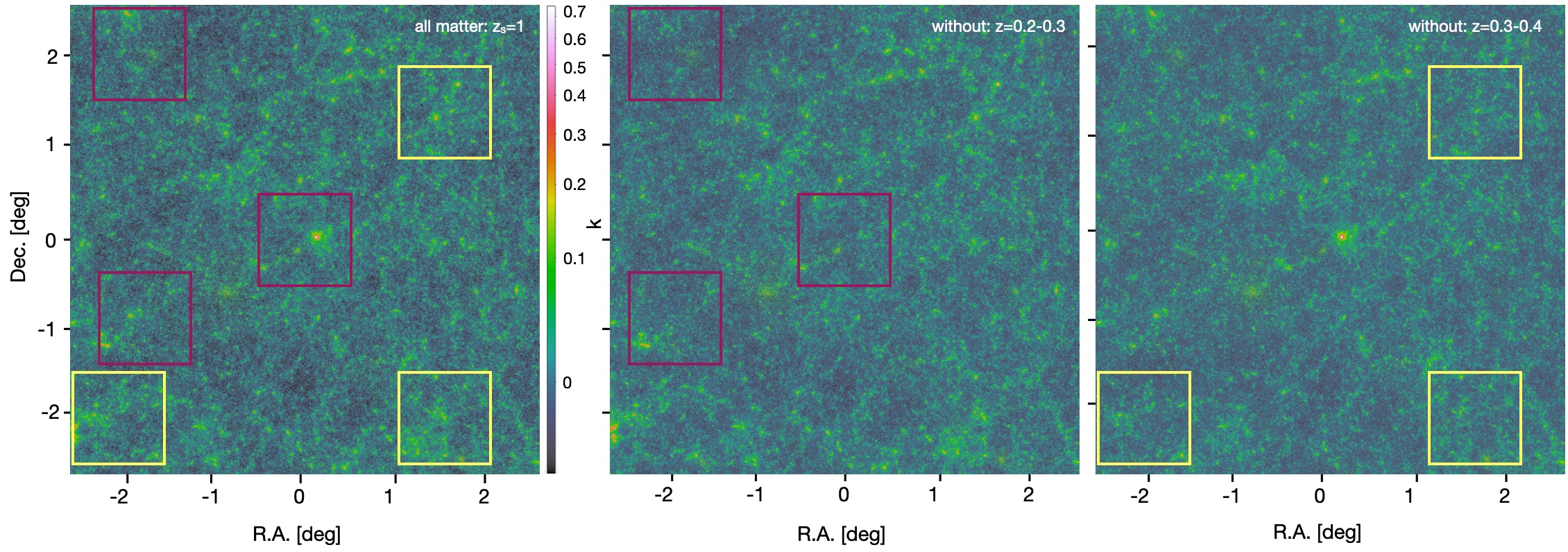}
\caption{Convergence map (5 deg on a side) for fixed source redshift $z_s=1$. 
In the left panel, we consider all the matter present along the line-of-sight from $z=0$ to $z=1$, while in the central and right panels, we remove the matter located in a redshift slice as indicated in the legend. The red and yellow squares highlight some particular high-density regions that disappear in the central and right panels when discarding the matter present in the considered redshift slices.}
\label{fig:conv_map_zs1}
\end{figure*}

\section{Weak lensing simulations}\label{WLsim}

To evaluate the performance of our methodology in identifying galaxy clusters via weak gravitational lensing (see the next section), we use dedicated light-cone simulations.

We build mass density planes piling the boxes of a cosmological simulation in a light-cone using the \texttt{MapSim} algorithm \citep{giocoli15} and shot the rays to construct lensing maps. This method was already tested and used on a variety of cosmological simulations \citep{tessore15,castro17,giocoli18a} and recently compared and validated with other algorithms \citep{hilbert20}, finding only per cent level differences for both cosmic shear two-point and peak statistics. We employ the \texttt{MapSim} pipeline to read the snapshots from an N-body simulation realisation and build a light-cone up to a redshift $z=4$. For the N-body run we make use of the reference $\mathrm{\Lambda}$CDM simulation of the DUSTGRAIN-{\em pathfinder} suite \citep{giocoli18b} which considers a volume of $(750$ $h^{-1}\,$Mpc$)^3$ filled with $N_{\rm p} = 768^3$ particles, corresponding to a particle mass resolution $m_{\rm p}=8 \times 10^{10} ~ h^{-1}\,M_{\odot}$. The initial conditions have been generated considering cosmological parameters consistent with the Planck-2015  data analysis \citep{planck1_15}, namely total matter density $\Omega_{\rm m}=0.31345$, baryon density $\Omega_{\rm b}=0.0491$, Hubble constant $H_0=67.31\,\mathrm{km\,s^{-1}\,Mpc^{-1}}$, scalar spectral index $n_{\rm s}=0.9658$, and mean amplitude of linear density fluctuations on the $8\,h^{-1}\,$Mpc scale $\sigma_8=0.842$. In particular, we use $21$ snapshots in the redshift range between $z=0$ and $z=4$. Given the box length of $750$ $h^{-1}$Mpc, $7$ boxes are needed to cover the comoving distance of about $5$ $h^{-1}\,$Gpc to a source redshift $z_{\rm s}=4$. The volume required to construct the light-cone is divided along the line-of-sight into multiple contiguous redshift slices, combining the individual snapshots to obtain better redshift sampling. If the redshift slice extends beyond the boundary of a single box, two lens planes are constructed from a single snapshot. The total number of lens planes up to $z_{\rm s} = 4$ is $27$. To avoid replicating the same structure along the line-of-sight, the periodic boxes are randomised by redefining the centre of the box and changing the ordering and signs of the axes \citep{roncarelli07}. From the line-of-sight plane realisations, we then shoot the rays using the Born approximation and construct convergence and shear maps on a square patch of $5\times5$ deg$^2$ for different source redshifts, considering a pixel resolution $2024 \times 2024$ on a grid.

In Fig.~\ref{fig:conv_map_zs1}, we show the convergence map obtained assuming a source redshift $z_s=1$. The left panel displays the convergence map obtained considering all matter along the line-of-sight, while in the central and right panels, we show the same map where we discard the matter distribution present between $z=0.2$--$0.3$ and $z=0.3$--$0.4$, respectively. The coloured squares highlight some particular features that are not present in the central and right panels. In particular, the red squares are related to the ones missing in the $z=0.2$--$0.3$ plane, while the yellow squares are the ones missing between $z=0.3$--$0.4$. The most noticeable case is the central cluster that is located between $z=0.2$--$0.3$ and is not present in the middle panel. This shows that, by removing individual planes one at a time, we can, therefore, use a `blinking' approach to gain some information about the redshift: this will be useful when matching dark matter haloes with weak lensing detections.

\begin{figure}
   \centering
   \includegraphics[width=\hsize]{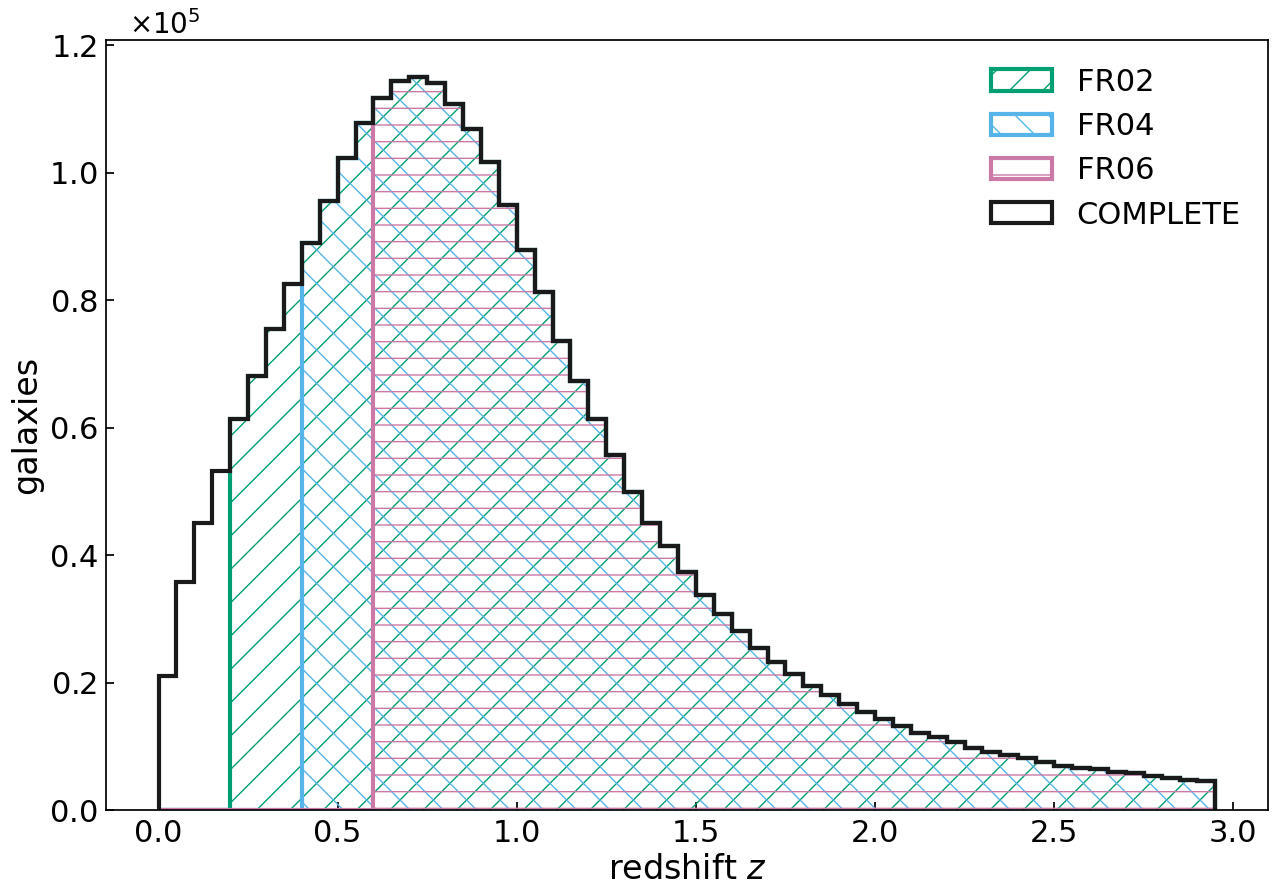}
      \caption{Redshift distribution of the weak lensing simulated sources. The black empty histogram refers to the COMPLETE catalogue, while the hatched histograms represent the foreground-removed catalogues FR02, FR04 and FR06 (described in Sect.~\ref{Filters}), shown as green, light-blue and magenta, respectively.
              }
         \label{fig:z_dist}
\end{figure}

\begin{figure*}
\centering
\includegraphics[width=\hsize]{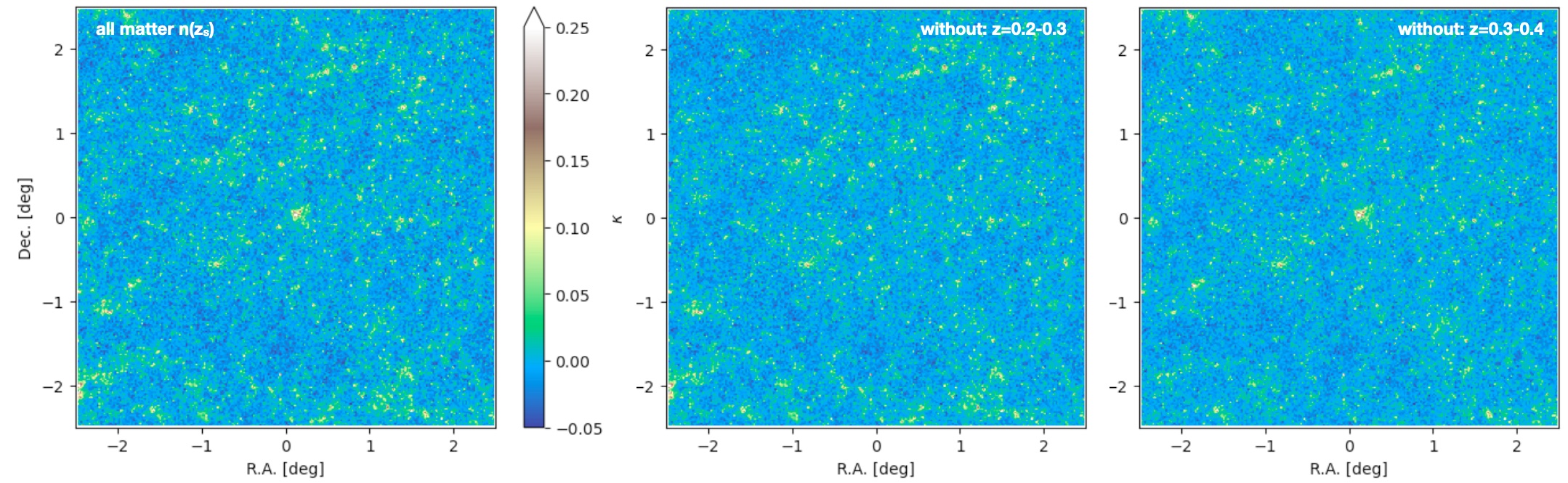}
\caption{Reconstructed convergence map from the source redshift distribution catalogue. The left panel refers to the full light-cone up to $z=4$, while in the central and right panels, we show the maps where we exclude the matter located between $z=0.2-0.3$ and $z=0.4-0.5$, respectively.}
\label{fig:conv_minus_planes}
\end{figure*}

We then populate the light-cone using a Euclid-like source redshift distribution (see Fig.~\ref{fig:z_dist}); we assume a distribution normalised to the total value of 30
galaxies per square arcmin, with a peak around $z=1$ \citep{boldrin12,boldrin16,giocoli24}, and we randomly distribute them in the map, neglecting source clustering. This gives us the possibility to construct a shear catalogue that will be the input for our \texttt{AMICO-WL} code. In computing the lensing quantities, we interpolate between two consecutive shear planes. To display the impact of the considered source redshift distribution on the lensing quantities, in Fig.~\ref{fig:conv_minus_planes}, we show the resulting convergence map. In the left panel, we consider all matter between $z=0$ and $z=4$, while in the central and right panels, we remove matter between $z=0.2$--$0.3$ and $z=0.4$--$0.5$, respectively. The effect of the missing structures in the central and right panels is, in this case, modulated by the source redshift distribution.

We also include the expected shear noise in the simulation to mimic realistic surveys. Uncertainties in the shear -- referred to as shape noise --  arise from a combination of the intrinsic shape of the galaxies and measurement errors, including, among other factors, uncertainties in the galaxy shape estimation and the point spread function correction. With the intrinsic shape of the galaxies being the dominant component and the galaxies being randomly distributed, the shape noise can be modelled as additive noise. This can be well approximated by a Gaussian distribution with a mean of $\mu=0$ and a standard deviation of $\sigma_{\epsilon}=0.26$ (e.g. \citealp{leauthaud2007weak}; \citealp{schrabback2015cfhtlens}, \citeyear{schrabback2018precise}). For each galaxy, the shape noise was, therefore, included in the catalogue by adding Gaussian noise to the two components of the shear. The final catalogue of galaxies is composed of the angular positions (i.e. right ascension and declination), the two perturbed components of the shear, and the true redshift -- which means that we do not consider any photometric redshift uncertainty in our analysis.

Our \texttt{MapSim} pipeline complements the particle light-cone with the line-of-sight 
corresponding halo and subhalo catalogues that were created at each simulation snapshot by applying a friend-of-friends algorithm (\citealp[]{davis1985evolution}) to identify the collapsed dark matter haloes. The algorithm used a linking distance $\lambda=0.16\times d$, where $d$ represents the mean separation distance between particles. Subsequently, the \texttt{SUBFIND} algorithm was applied to determine for each friend-of-friends-identified halo the mass $M_{200}$ and the radius $R_{200}$, which correspond to the mass and the radius of a spherical region around the fiducial centre of each halo, enclosing 200 times the critical density of the Universe, and the corresponding subhalo population. The resulting catalogue comprises the angular positions (i.e., right ascension and declination) of the identified dark matter haloes and subhaloes, along with their estimated redshift ($z_l$), mass, and radius. This is the catalogue used in the matching procedure to test the performance of the \texttt{AMICO-WL} algorithm (see Sect.~\ref{LPmatch}).


\section{Optimal filtering technique}\label{OptFilt}

In this section, we describe the technique on which \texttt{AMICO-WL} is based, the optimal matched filtering, and a linear weak-lensing estimator. 

\subsection{Definition of the optimal filter}
The optimal matched filter implemented in \texttt{AMICO-WL} was first presented in \citet{maturi2005optimal} and \citet{maturi2007searching} for an unbiased detection of the weak lensing signal of dark matter haloes aiming at a minimum variance in the estimates, achieved by suppressing the contamination of the large-scale structures and thus reducing the number of spurious detections. The performance of the weak lensing application of the optimal filter was then compared with other weak lensing estimators in \citet{pace2007testing}. In contrast to the standard implementation of AM, the shape of the optimal filter is determined not only by the shear profile of the lens but also by the correlation properties of the noise affecting the weak lensing measurements. Here, we model the contribution of virialised objects by assuming a radial profile for the shear signal and the properties of the LSS by the corresponding power spectrum. In particular, for galaxy clusters, we assume that the shear signal is produced by a Navarro-Frenk-White (NFW) profile (\citealp{navarro2004inner}; \citealp{navarro1996structure}). 

The weak gravitational lensing signal of a dark matter halo, $S(\boldsymbol{\theta})$, is given by an amplitude, $A$, and an angular shape, $\tau(\boldsymbol{\theta})$. The data we measure, $D(\boldsymbol{\theta})$, are contaminated by the noise, $N(\boldsymbol{\theta})$, so that they can be modelled as:
\begin{equation}
\label{eqn:data}
     D(\boldsymbol{\theta}) = S(\boldsymbol{\theta}) + N(\boldsymbol{\theta}) = A\tau(\boldsymbol{\theta}) + N(\boldsymbol{\theta}).
\end{equation}
The linear filter is defined as the convolution kernel, $\Psi(\boldsymbol{\theta})$,  applied to the data, $D(\boldsymbol{\theta})$, in order to yield an optimal estimate, $A_{est}$, of the amplitude of the signal at the position $\boldsymbol{\theta}$:
\begin{equation}
\label{eqn:Estim}
    A_{est}(\boldsymbol{\theta}) = \int_{\mathbb{R}^2} D(\boldsymbol{\theta}')\Psi(\boldsymbol{\theta - \theta'})\text{d}^2\boldsymbol{\theta'}.
\end{equation}
The filter used in this linear estimator must satisfy two constraints. First, $A_{est}$ must be unbiased, i.e. its average $b$ over many realisations has to vanish:
\begin{equation}
\label{eqn:bias}
    b \equiv \langle A_{est} - A \rangle = A \left[ \int_{\mathbb{R}^2}\tau(\boldsymbol{\theta'})\Psi(\boldsymbol{\theta'})\text{d}^2\boldsymbol{\theta'}\right] = 0.
\end{equation}
Second, the measurement noise $\sigma^2$, determined by the mean-squared deviation of the estimate from its true value, 
\begin{equation}
\label{eqn:varEst}
\begin{split}
    \sigma^2  = \langle(A_{est} - A)^2\rangle
= b^2 + \frac{1}{(2\pi)^2}\int_{\mathbb{R}^2} |\hat{\Psi}(\boldsymbol{k})|^2P_N(k)\text{d}^2\boldsymbol{k} 
\end{split}
\end{equation}
has to be minimal. To find the function $\Psi$ that satisfies these two conditions, we combine them by means of a Lagrangian multiplier $\lambda$, we carry out the variation of $L=\sigma^2+\lambda b$ with respect to $\Psi$ and thus find the filter $\Psi$ minimising $L$. The solution to this constrained minimisation in the Fourier domain is given by
\begin{equation}
\label{eqn:filtFunct}
    \hat{\Psi}(\boldsymbol{k}) = \frac{1}{(2\pi)^2}\left[ \int_{\mathbb{R}^2}\frac{|\hat{\tau}(\boldsymbol{k)}|^2}{P_N(k)}\text{d}^2\boldsymbol{k}\right]^{-1} \frac{\hat{\tau}(\boldsymbol{k})}{P_N(k)} \;,
\end{equation}
where $\hat{\tau}(\boldsymbol{k)}$ is the Fourier transform of the expected shear profile of an NFW halo and $P_N$ is the noise power spectrum. The filter is therefore constructed to be most sensitive for those spatial frequencies where the signal $\hat{\tau}$ is large and the noise $P_N$ is small. Finally, we back-Fourier transform $\hat{\Psi}(k)$ into the real space to perform the estimates.

For weak lensing observations, the signal can be either the convergence or the shear. However, since the convergence is not directly measurable, it is convenient to use the shear and, specifically, the tangential shear component. In this case Eq. (\ref{eqn:Estim}) takes the form
\begin{equation}
\label{eqn:EstimAp}
    A_{est}(\boldsymbol{\theta}) = \int_{\mathbb{R}^2} \gamma_{obs}(\boldsymbol{\theta'};\boldsymbol{\theta)}\Psi(\boldsymbol{\theta - \theta'})\text{d}^2\boldsymbol{\theta'} \; .
\end{equation}

 The noise is assumed to be given by three contributions, namely the contributions from the finite number of background sources, the intrinsic random ellipticities and orientations, and the correlated weak-lensing signal due to the large-scale structure of the Universe.
The first two are characterized by the power spectrum
\begin{equation}
\label{eqn:NPell}
    P_{\epsilon}=\frac{1}{2}\frac{\sigma^2_{\epsilon_s}}{n_g}\,,
\end{equation}
which depends on the dispersion of the intrinsic ellipticities of the sources $\sigma_{\epsilon_s}$, and on the number density of background galaxies $n_g$. 
The contribution from the large-scale structures can be described by the shear power spectrum $P_{\gamma}(l)$, 
thus the total noise power spectrum can be written as:
\begin{equation}
\label{eqn:totNP}
    P_N(k)=P_{\epsilon}(k)+P_{\gamma}(k) ,
\end{equation}
where $P_{\gamma}$ is computed assuming non-linear theory models \citep{camb,smith03,takahashi12}. 

\subsection{Filter implementation and foreground removal}\label{Filters}

In our analysis, we implemented four different realisations of the optimal filter:  together with the whole galaxy catalogue, we considered three redshift cuts, corresponding to all galaxies with redshift $z>z_{\rm min}$, with $z_{\rm min}=0.2$, $z_{\rm min}=0.4$ and $z_{\rm min}=0.6$. These redshift distributions are also shown in Fig.~\ref{fig:z_dist}. This foreground removal strategy is adopted in order to clean the data set from the noise that foreground galaxy ellipticities produce on the weak lensing measures on the lens planes located at a higher redshift. 
The redshift-cut shear catalogues allow us to select galaxies that are primarily background sources, thus serving as ideal targets for weak lensing detection of dark matter haloes. Galaxies below the redshift thresholds $z_{\rm min}$ could be background sources for some clusters but often act as foreground objects for the majority of clusters of interest. This foreground contamination can introduce noise into the weak lensing signal, hindering our ability to detect and accurately characterise dark matter haloes.

\begin{figure}
   \centering
   \includegraphics[width=\hsize]{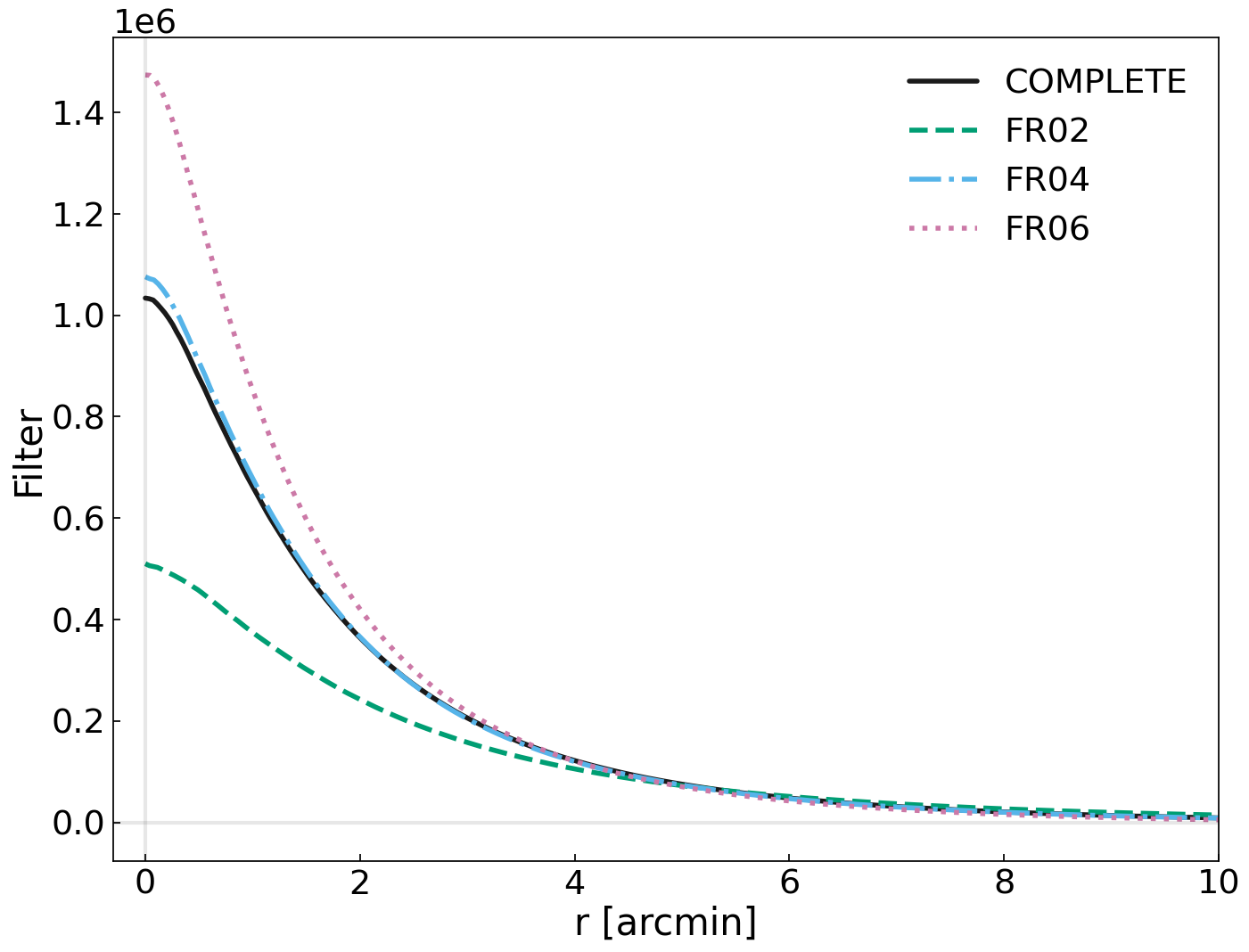}
      \caption{Radial profiles (in real space) of the optimal filters employed for the four \texttt{AMICO-WL} runs: COMPLETE (black solid line), FR02 (green dashed line), FR04 (cyan dash-dotted line) and FR06 (pink dotted line).}
         \label{fig:filters}
\end{figure}

The optimal filter must be optimised according to the specific properties of each truncated catalogue. For this purpose, the cosmic shear power spectrum and the effective density of galaxies have to be adopted. In Table \ref{tab:filters}, we summarize the properties of the four filters.
For the template halo, we assume a standard set of parameters already used, for instance, in \citet{maturi2007searching}. For the COMPLETE catalogue, the parameters of the template halo are meant to detect high-mass haloes that have an intermediate distance between the observer and the sources, a condition for which the lensing effect is maximised due to the geometrical dependencies of the lensing strength \citep{pace2007testing}. For the filters applied to the foreground removed catalogues, the template lens redshift is set instead to $z_{\rm min}$, the redshift value of the threshold used for the cut. In Fig.~\ref{fig:filters}, we show the profile of the four filters in real space. From the figure, it is clear that the lower the redshift of the template model, the shallower the central peak of the filter, i.e. the broader the filter. This is clear by noting the similarity in the profile of the COMPLETE and FR04 optimal filters, where the lens redshift is set to $z_l=0.4$.

\begin{table*}
\caption{Parameters of the filters for the four catalogues of simulated galaxy ellipticities.}
\centering
\def\arraystretch{1.3}
\begin{tabular}{lccccc}
\hline\hline
Parameter&  &COMPLETE&FR02&FR04&FR06\\

\hline
 \small{{lens mass}}  & {$M_l$ [$h^{-1}M_{\odot}$]} & {$1\times10^{15}$}   & {$1\times10^{15}$}  & {$1\times10^{15}$}& {$1\times10^{15}$} \\ 
 \small{{lens redshift}}   & {$z_l$} & 0.4  & 0.2 & 0.4 & 0.6 \\ 
\small{{source redshift}}  & {$z_s$} &  0.93 & 0.98 & 1.06 & 1.18  \\ 
\small{{white noise}} & {$N_{\epsilon}$} & {$1.27\times10^{-10}$}  & {$1.35\times10^{-10}$} & {$1.52\times10^{-10}$} & {$1.84\times10^{-10}$}  \\ 
\small{{beam FWHM }} & {$b_{FWHM}$ [arcmin]} &  0.182  & 0.188 & 0.199 & 0.220     \\ 
\hline
\end{tabular}
\label{tab:filters}
\end{table*}
From here on, we will refer to the application of the first optimal filter shown in Table~\ref{tab:filters} to the catalogue of galaxy ellipticity without cuts as COMPLETE, while the application of the other three optimal filters with lens redshift cut equal to $z_{\rm min}=0.2$, $z_{\rm min}=0.4$ and $z_{\rm min}=0.6$ to the correspondent foreground removed sub-catalogue as FR02, FR04 and FR06, respectively.  

\section{\texttt{AMICO-WL} algorithm}\label{AMICOWLapp}

The \texttt{AMICO} algorithm (\citealp{bellagamba2011optimal}; \citealp{bellagamba2018amico}) was developed to identify galaxy clusters using photometric galaxy catalogues. After a dedicated challenge presented in \citet{adam19}, it has been selected by the Euclid Collaboration as one of the algorithms that will be used to produce the official Euclid galaxy cluster catalogues. 
The original version of the code, assuming a model for the density profile and for the luminosity function of galaxies in clusters, identifies galaxy groups and clusters by finding overdensities and matching them with the defined model based on the optimal filter technique, as presented in Sect.~\ref{OptFilt}. The versatility of Optimal Matched Filtering offers the possibility to implement other observables, as, in our case, the ones based on weak-gravitational lensing. 

In this work, we modified \texttt{AMICO} to process the gravitational lensing signal of galaxy clusters derived from the ellipticity measurements of background galaxies.  We implemented a new weak-lensing branch in \texttt{AMICO} by exploiting the polymorphism of \texttt{C++} to minimise the modifications of the well-tested existing code. We call this modified version \texttt{AMICO-WL}. 
Our algorithm follows the layout of the original code, which can be broken down into two main steps:
\begin{enumerate}

    
    \item \textbf{Map creation}: the optimal filter technique is applied to the data, producing amplitude and variance maps of the signal.
    
    \item \textbf{Detection}: the $SNR$ map is computed, and the clusters are detected as peaks, also employing a \textit{cleaning} procedure.
\end{enumerate}
In this section, we describe \texttt{AMICO-WL} in detail, first explaining in Sect.~\ref{MapCre} the method used to create maps and then describing in Sect.~\ref{SNR} the detection algorithm, including the implemented cleaning procedure and in Sect.~\ref{Clean}  the method for the computation of the optimal $SNR$ detection threshold.

\subsection{Signal map creation}\label{MapCre}

In order to compute the signal maps, the algorithm needs a catalogue of galaxies reporting their positions (cartesian or sky coordinates), the two components of the ellipticity, the redshift $z_g$, necessary for the foreground removal, and their corresponding weak lensing weight $w$ (if available).

The algorithm identifies all galaxies that fall in each pixel, then computes the lensing contribution related to those galaxies in the surrounding region according to the filter value and finally adds to the amplitude map, taking into account the weight provided by the filter $\Psi$.
This is done by replacing the integral in Eq. (\ref{eqn:Estim}) with a sum over all galaxies, computed using the tangential component of the ellipticity $\epsilon_+$, which is a good estimator of $\gamma_+$ (see Sect. \ref{WL_ell}), as
\begin{equation}
\label{eqn:EmodeMap}
   E_{map}(\boldsymbol{\theta}) \equiv A_{est}(\boldsymbol{\theta}) 
   = \frac{1}{n_g}\sum_{k}\epsilon_+(\boldsymbol{\theta};\boldsymbol{\theta_k})w_k\Psi(|\boldsymbol{\theta_k}-\boldsymbol{\theta}|) \;,
\end{equation}
where the sum is extended to all galaxies within a circular area centred on $\vec{\theta}$ and extending up to a cut radius, which is a user-defined parameter. If the lensing weights $w$ are not used, they are internally set to unity. The normalisation factor $n_g$ is the effective number density of galaxies computed as the sum of the lensing normalised statistical weights of the galaxies.
Along the amplitude map, the code produces the variance map,
that can be written as:
\begin{equation}
\label{eqn:varE}
\sigma_{E_{mode}}^2(\boldsymbol{\theta})\equiv\sigma_{A_{est}}^2(\boldsymbol{\theta})=\frac{1}{2n_g^2}\sum_{k}|\epsilon_+(\boldsymbol{\theta};\boldsymbol{\theta_k})|^2w_k^2\Psi^2(|\boldsymbol{\theta_k}-\boldsymbol{\theta}|) \;.
\end{equation}
Note that this noise estimate only accounts for the galaxy-shape noise. To derive the total noise, we add in quadrature the LSS noise contribution given by:
\begin{equation}
\label{eqn:CLSS}
    \sigma^2_{LSS}=\frac{1}{2\pi}\int_0^{\infty}|\hat{\Psi}(\boldsymbol{k})|^2P_{\gamma}(k)k\text{d}k \;.
\end{equation}

The $B$-mode is evaluated by exploiting the same algorithmic infrastructure by simply replacing the ellipticity $\epsilon_+$ in Eq.~(\ref{eqn:EmodeMap}) with the cross component of the ellipticity $\epsilon_{\times}$. This $B$-mode can be used to estimate the noise fluctuations and possible systematics in the shape measurement pipeline, as they do not contain any lensing contribution by construction (see Eq. \ref{eqn:gammaCrossAve}).
%


%
\begin{figure*}
\begin{tabular}{c}
\resizebox{\hsize}{!}{\includegraphics{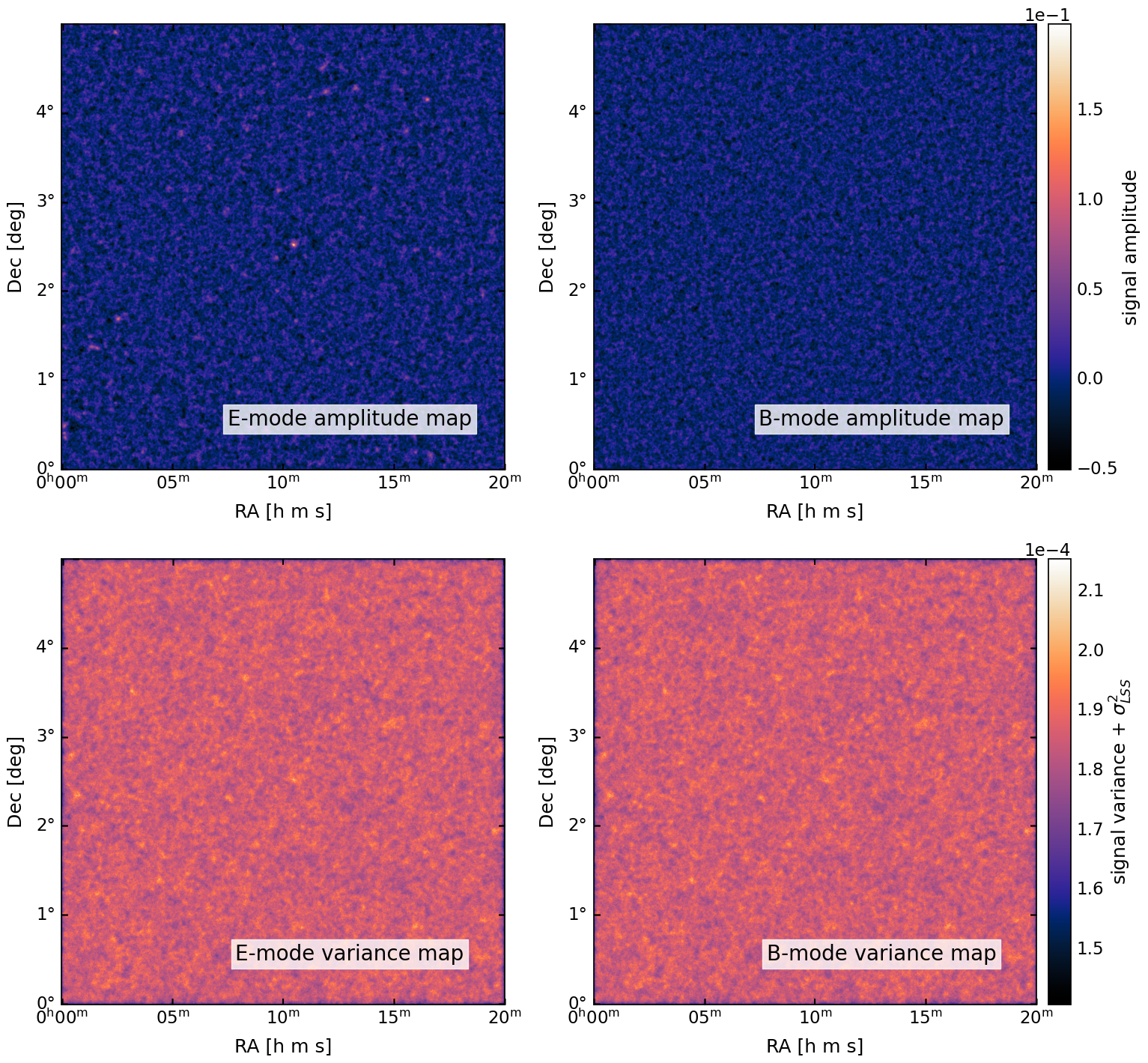}}
\end{tabular}
\caption{The $E$-mode amplitude and its variance lensing signal maps are shown in the top-left and bottom-left panels, respectively, while the $B$-mode amplitude and its variance lensing signal maps are shown in the top-right and bottom-right panels. The maps cover a field of view  5° $\times$ 5° and refer to the COMPLETE run.
}
\label{fig:maps}
\end{figure*}
In Fig.~\ref{fig:maps}, we show the $E$-mode amplitude (top-left panel) and the corresponding variance map (bottom-left panel) returned by \texttt{AMICO-WL} in the total 5° $\times$ 5° field of view for the COMPLETE case, which considers all galaxies with no redshift cuts. 
The haloes with the largest signal are visible as clear peaks in the top panel. Comparing the amplitude map obtained by \texttt{AMICO-WL} with the corresponding simulated convergence map (Fig.~\ref{fig:conv_map_zs1}) we notice that the largest peaks in the $E$-mode map clearly match the structures of the convergence map. An accurate definition of the detected peaks and the matching with the structures in the simulations will be discussed in Sect.~\ref{LPmatch}. In Fig.~\ref{fig:maps} we also show the $B$-mode amplitude (top right panel) and variance maps (bottom right panel). The lensing signal peaks present in the $E$-mode cannot be found in the $B$-mode signal map, which is only produced by noise and thus appears as the background map of the $E$-mode.

\subsection{SNR threshold selection} \label{SNR}

Once the amplitude and variance maps have been created, we proceed with the detection of the peaks. The detection is not carried out on the amplitude map but on the $SNR$ map, which \texttt{AMICO-WL} computes on each pixel as the ratio between the amplitude value and the square root of the median value of the variance map. \texttt{AMICO-WL} also provides the possibility to compute $SNR$ locally, i.e., by computing the noise using the variance map value in the pixel instead of the median value. Since the variance is highly homogeneous, as shown in the bottom left panel of Fig.~\ref{fig:maps}, using the non-local value of $SNR$ improves the stability of the detection procedure. However, this approach is reliable whenever the uncertainties are homogeneous on large scales.

\begin{figure}
   \centering
   \includegraphics[width=\hsize]{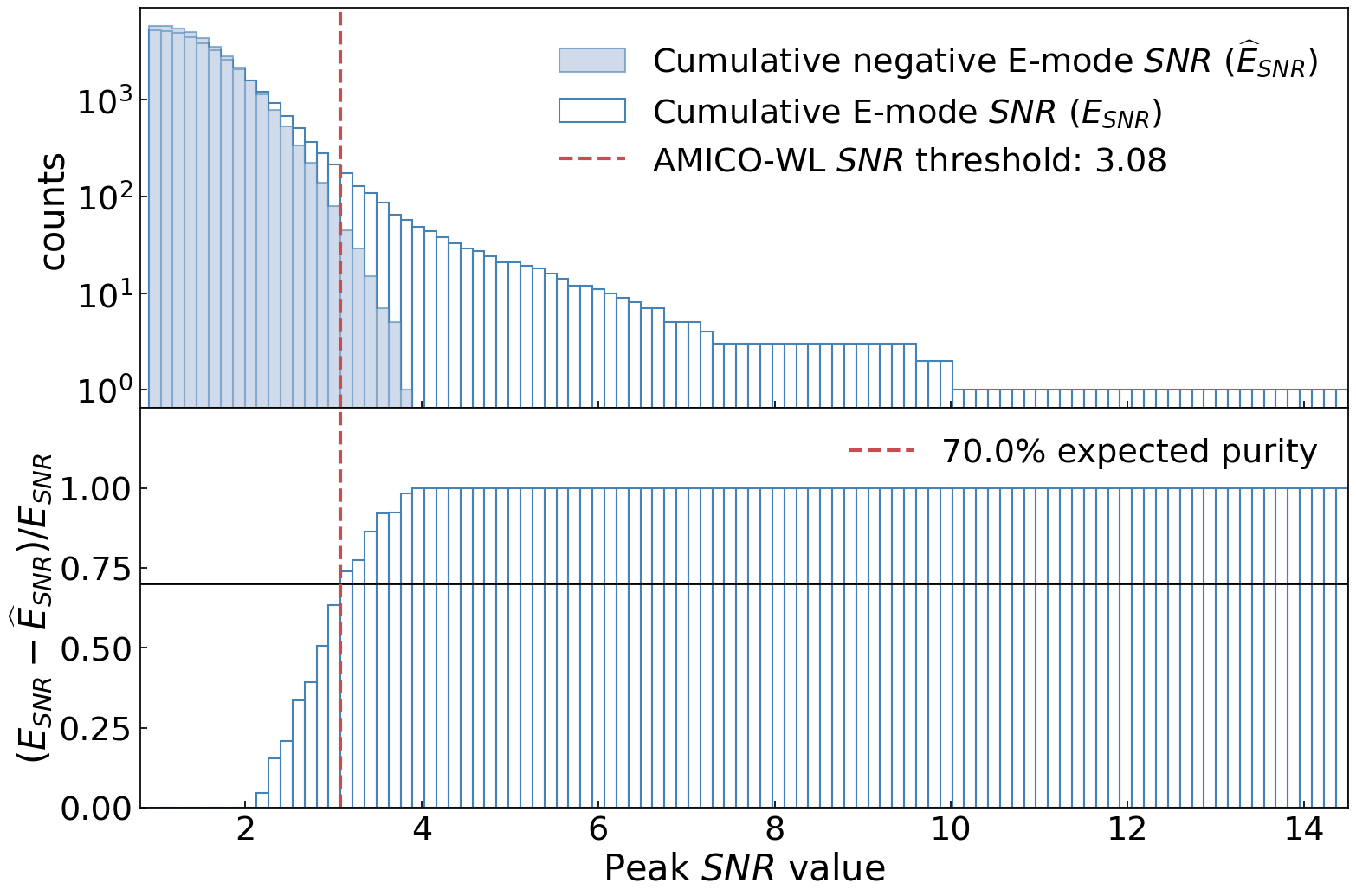}
      \caption{$SNR$ threshold determination analysis.
      Top panel: cumulative $SNR$ distributions of positive ($E_{SNR}$) and absolute negative peaks ($\hat{E}_{SNR}$) in the $E$-mode map detected with SExtractor. 
      Bottom panel: difference in cumulative counts of positive and negative peaks as a function of the $SNR$. The $SNR$ threshold achieving a 70\% purity is defined where the cumulative positive peak count reaches 70\% of the total (horizontal line) and is marked across both panels (red vertical dashed line).}
         \label{fig:peak_statistics}
\end{figure}

To determine the minimum $SNR$ used to define a reliable sample of detections, we developed a method that exploits the statistical properties of the positive and negative $E$-mode peaks in these maps. As a technical note, we use SExtractor \citep{bertin1996sextractor} to detect the peaks in the maps down to very low $SNR$, and when looking for the negative peaks we simply invert the sign of the map. The positive peaks in the $E$-mode map are generated by ($i$) lensing produced by galaxy clusters, ($ii$) LSS overdensities or ($iii$) noise. In contrast, the negative minima of the $E$-mode map are generated by ($i$) noise and ($ii$) lensing signal from LSS underdensities\footnote{With good approximation, the lensing signal of the LSS is well described by a Gaussian random field where negative peaks follow the same statistics of the positive ones.} but not by galaxy clusters. We exploit this key difference between the positive and negative peaks to evaluate the sample purity by comparing the $SNR$ cumulative distribution obtained from the original $E$-mode amplitude map ($E_{SNR}$) and that of the absolute values of the negative peaks ($\hat{E}_{SNR}$). The SNR cumulative distribution of the peaks for the COMPLETE run is shown in the upper panel of Fig.~\ref{fig:peak_statistics}.

As the statistics of the negative peaks reflect the presence of spurious detections due to noise and the LSS contribution, as already discussed, we use it to define the sample purity as:
\begin{equation}
    \mathcal{P}=\frac{E_{SNR}-\hat{E}_{SNR}}{E_{SNR}}.
\end{equation}
The purity values as a function of SNR are shown in the bottom panel of Fig.~\ref{fig:peak_statistics}: according to the desired value of purity from this distribution 
it is possible to select a $SNR$ threshold that \texttt{AMICO-WL} uses as the lower detection threshold.
For example, a sample purity of 70\% is expected to be obtained with a threshold of $SNR=3.08$ (shown by the vertical red line in the plot).
\subsection{Cleaning method}\label{Clean}

The algorithm starts by identifying the pixel with the largest $SNR$ and stores a unique identification number (ID) for this detection, the pixel position (Xpix, Ypix), the sky position (RA and Dec) in degrees $\boldsymbol{\theta}_d$, the signal-to-noise ratio ($SNR$) and the associated value of the amplitude map ($A$). Then, it proceeds by subtracting the signal contribution of this detection from the amplitude map to facilitate the detection of other possible nearby contributors. 
This is done by removing the filtered lensing signal carried by each individual galaxy that we model using the same shear template adopted for the filter construction.
Given a detection centred at the position $\boldsymbol{\theta}_d$, the expected shear modulus at the location $\boldsymbol{\theta}$ is
\begin{equation}
\label{eqn:shearMod}
    |\gamma_d(\boldsymbol{\theta};\boldsymbol{\theta}_d)| = A \cdot \tau(\boldsymbol{\theta};\boldsymbol{\theta}_d) \;,
\end{equation}
where the modelled shear profile $\tau(\boldsymbol{\theta})$ is scaled by the detection amplitude $A$ associated to the detection. The resulting complex shear attributed to the detection at each position is thus
\begin{equation}
\label{eqn:gammaDet}
    \gamma_d = \gamma_{d,1} + i\gamma_{d,2} = |\gamma_d|(\cos{2\varphi}+i\sin{2\varphi})\;,
\end{equation}
which is tangential with respect to the detection location by construction.

Based on that, the two components of the tangential shear are:
\begin{equation}
\label{eqn:gammaDetComps}
    \begin{aligned}
        \gamma_{d,1}(\boldsymbol{\theta};\Delta\theta) &= - |\gamma_d(\boldsymbol{\theta};\Delta\theta)|\cdot\cos{\left(2\arctan{\frac{\Delta\theta_1}{\Delta\theta_2}}\right)}\; \\
        \gamma_{d,2}(\boldsymbol{\theta};\Delta\theta) &= - |\gamma_d(\boldsymbol{\theta};\Delta\theta)|\cdot\sin{\left(2\arctan{\frac{\Delta\theta_1}{\Delta\theta_2}}\right)}
    \end{aligned}
    \; ,
\end{equation}
where, in the flat sky approximation, we can write $\Delta\theta = \theta - \theta_d = \Delta\theta_1 +i\Delta\theta_2$, with $\Delta\theta_1$ and $\Delta\theta_2$ being the difference in position of the galaxy and detection with respect to the two cartesian coordinates.
With this as the new shear components of the galaxy, the code computes the lensing contribution and removes it from the pixels of the map in the region surrounding the detections.

After the removal of the detection contribution, that is, by applying the described procedure to all galaxies, the $SNR$ is re-evaluated, and the algorithm continues by considering the pixel with the next largest $SNR$, reiterating the previous process. It may happen that, after cleaning, some peaks that initially had a $SNR$ value below the detection threshold can be boosted to be detectable. Therefore, detections are not identified with a strict monotonic decrease in their value of $SNR$. 

\begin{figure}
\begin{tabular}{c}
\resizebox{\hsize}{!}{\includegraphics{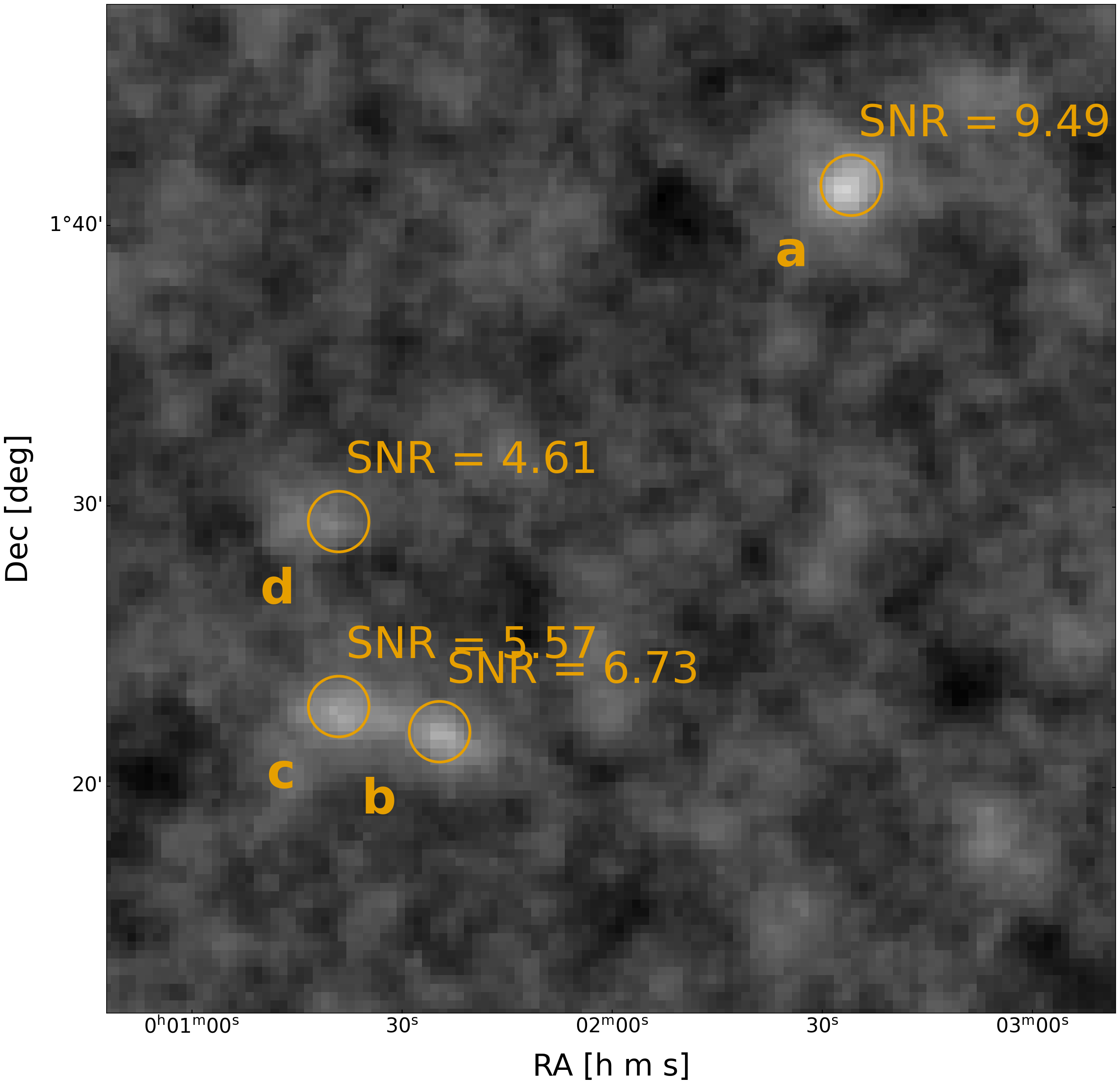}} \\
\resizebox{\hsize}{!}{\includegraphics{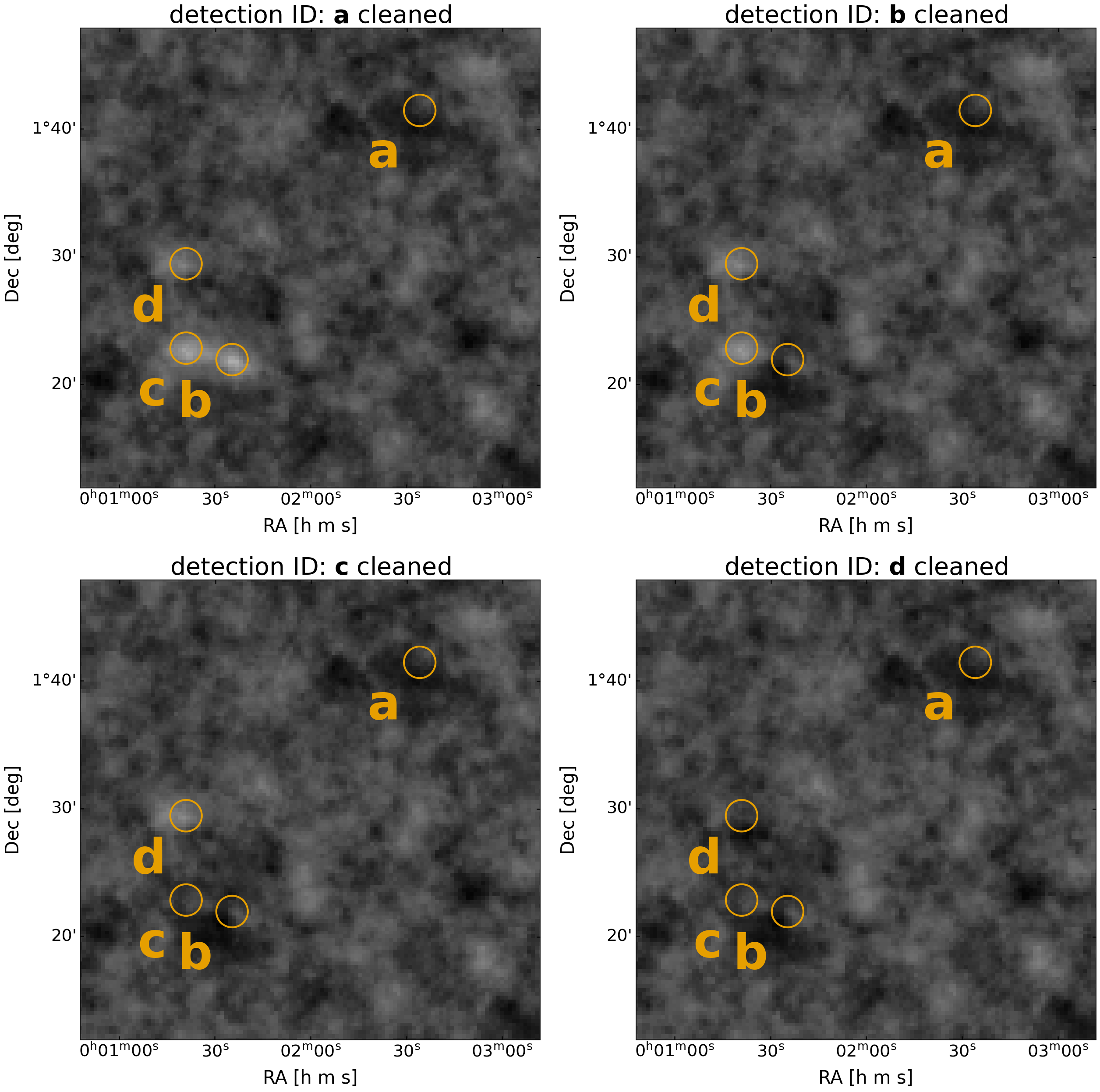}} \\
\end{tabular}
\caption{Example of the cleaning procedure on a 0.36 deg$^2$ region containing 4 signal peaks, indicated as a, b, c and d. Each circle is centred on the pixel with the maximum amplitude. The corresponding $SNR$ is also reported.  Top panel: the amplitude map before cleaning. 
Smaller bottom panels: the same after each of the sequential cleaning steps. The algorithm detects and removes each peak in descending $SNR$ order: first peak a (top-left), then peak b (top-right), peak c (bottom-left), and finally peak d (bottom-right), leaving at the end a map without any significant signal.
}
\label{fig:cleaning}
\end{figure}

A visual impression of the cleaning procedure is shown in Fig.~\ref{fig:cleaning}, where four peaks are progressively detected and removed with the discussed procedure. 
When the algorithm reaches the minimum $SNR$ threshold imposed by the user (for instance, the one based on the method described in Sect.~\ref{SNR}), 
the procedure ends, and the iteration stops. Together with the catalogue of detections, the code outputs the amplitude map where all the lensing signals from the dark matter halo candidates identified during the run have been removed. 

\section{Results and statistical analysis}\label{StaAna}

We now describe the results of the application of the \texttt{AMICO-WL} algorithm with the four filters discussed in Sect.~\ref{Filters} to the simulated catalogue of galaxies ellipticities described in Sect.~\ref{WLsim}, with the four different redshift cut-offs. From a visual inspection, the resulting $E$-mode, $B$-mode, and variance maps of these four cases appear similar, but the differences cannot be neglected, especially in the low $SNR$ tail, because the application of the filters shown in Fig.~\ref{fig:filters} produces
maps with different values. Moreover, the normalisation of the maps changes as the effective number density of sources $n_g$ is different for the four catalogues of galaxies: cutting out foreground galaxies leads to a decrease in the surface number density from $n_g=30.0$ gal/arcmin$^2$ for COMPLETE to $n_g=20.7$ gal/arcmin$^2$ for FR06. Last but not least, a significant variation in the variance map comes from the change of $\sigma^2_{LSS}$ that depends on the specific filter $\Psi$, and on the cosmic shear power spectrum $P_{\gamma}$, which is related to the redshift distribution of sources (see Eq.~\ref{eqn:CLSS}). 

For the four applications, we report in the upper part of Table~\ref{tab:PUR_COMP} the resulting $SNR$ thresholds corresponding to an expected purity of 70\% as obtained with the method described in Sect.~\ref{Clean}. The table also shows the corresponding number of detections. 

\begin{figure}
   \centering
   \includegraphics[width=\hsize]{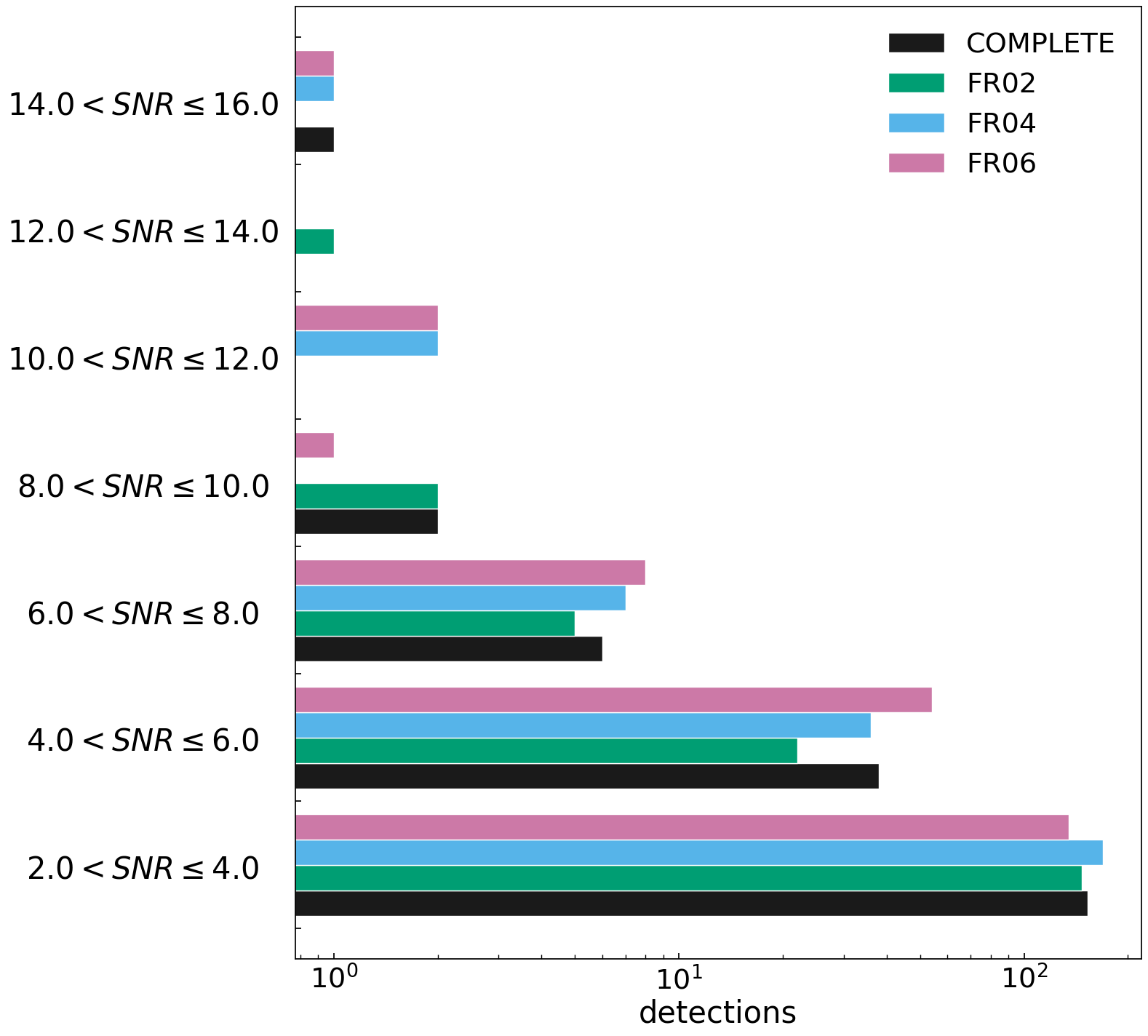}
      \caption{Distribution of signal-to-noise ratio $SNR$ in the catalogues of detections in $\Delta SNR=2.0$ bins: COMPLETE detections in black, FR02 in green, FR04 in light blue and FR06 in pink bars. The counts are on a logarithmic scale.}
        \label{fig:hist_snr}
\end{figure}

In Fig.~\ref{fig:hist_snr} we show the $SNR$ distribution of the detections. In the four cases, most of the detections have $SNR<4.0$, namely 76\% for COMPLETE, 83\% for FR02, 79\% for FR04 and 67\% for FR06. In the high-end $SNR$ tail, we found the same peaks in all four realisations but with different values of $SNR$.

%

%

%
%

\begin{table*}
\caption{Summary of the detection methods performance and statistical analysis}
\centering
\def\arraystretch{1.2}
\begin{tabular}{llcccc}
\hline\hline
& & COMPLETE  & FR02 & FR04 & FR06 \\
\hline
 \multirow{4}{3em}{TOTAL} & $SNR$ threshold & 3.08 & 2.69 & 3.07 & 3.34 \\  
  & \# detections & 200 & 177 & 215 & 201 \\
  & COMP [\%] & 13.0 & 13.0 & 14.0 & 13.0  \\
  & PUR [\%] & 61.0 & 71.0 & 64.0 & 61.0  \\
\hline
 \multirow{3}{3em}{$\sim$70\% PUR} & $SNR$ threshold & 3.50 & 2.69 & 3.25 & 3.54 \\  
  & \# detections & 90 & 177 & 152 & 135 \\
  & COMP [\%] & 6.5 & 13.0 & 11.1 & 9.9  \\
\hline
 \multirow{3}{3em}{$\sim$80\% PUR} & $SNR$ threshold & 3.66 & 3.20 & 3.72 & 4.05 \\  
  & \# detections & 64 & 70 & 67 & 61 \\
  & COMP [\%] & 5.3 & 5.8 & 5.5 & 5.1  \\
\hline
\end{tabular}
\label{tab:PUR_COMP}
\end{table*}

\subsection{Lens plane matching procedure}\label{LPmatch}

To assess the performance of \texttt{AMICO-WL} and compare the results obtained with the four different implementations of the optimal filter, it is necessary to cross-match the detections with the haloes in the simulations and distinguish the matched detections from those caused by noise fluctuations and contribution due to the large-scale structures. The most common procedure to perform this matching is to use a fixed physical, comoving, or angular radius centred on the detection and associate it to the closest halo inside the radius \citep[see e.g.][]{hennawi2005shear,gavazzi2007weak,miyazaki2018large,hamana2020weak,oguri2021hundreds}. 
Even using an adaptive matching radius, the method is afflicted by the lack of redshift information on the weak-gravitational lensing side, leading to a high probability of including random matches due to the large number of haloes displaced along the line-of-sight of a detection. Moreover, this approach does not allow us to identify the actual halo responsible for the largest lensing contribution (i.e. the one causing the detection) and does not allow us to distinguish the cases in which a detection is caused by a series of small haloes displaced along the line-of-sight, which would not be detected individually.
 
For these reasons, we followed a method similar to \citet{pace2007testing} to gain some redshift information regarding the weak lensing detections when performing their cross-matching with the haloes in the simulations. This approach consists of a `blinking'" procedure exploiting the fact that haloes are well-localised structures, i.e. their signal originates from a single thin lens plane.
In contrast, observational noise, uncorrelated structures, and projection effects originate from all redshift planes. Our matching method relies on the application of \texttt{AMICO-WL} to 10 different realisations built from the same simulation, but in which we removed from the ray-tracing the contribution of all haloes contained in redshift slices of $\Delta z=0.1$, one at a time (see Sect.~\ref{WLsim} for a discussion on the effect on the convergence maps and
Figs.~\ref{fig:conv_map_zs1} and ~\ref{fig:conv_minus_planes} for a visual impression). The thickness of the redshift slices is chosen for practical reasons: they are sufficiently thin for the purpose but large enough to keep the computational costs small. In this way, the lensing signal of a detection caused by a halo contained in the removed redshift slice is completely removed.
In contrast, false detections due to noise, uncorrelated structures, or projection effects are only marginally affected by the removal, as their contribution comes from more redshift planes. This allows us to gain information about the redshift of detections, thus improving the reliability of the cross-matching with the halo catalogue.


\citet{pace2007testing} first matched the detections with the catalogue of haloes (over the entire redshift range) with a simple positional match and then removed the redshift plane that contained the matched halo. If the $SNR$ drops by more than 25\%, the candidate is considered to be connected with the single dark matter halo in the plane. 

\begin{figure}
   \centering\includegraphics[width=\hsize]{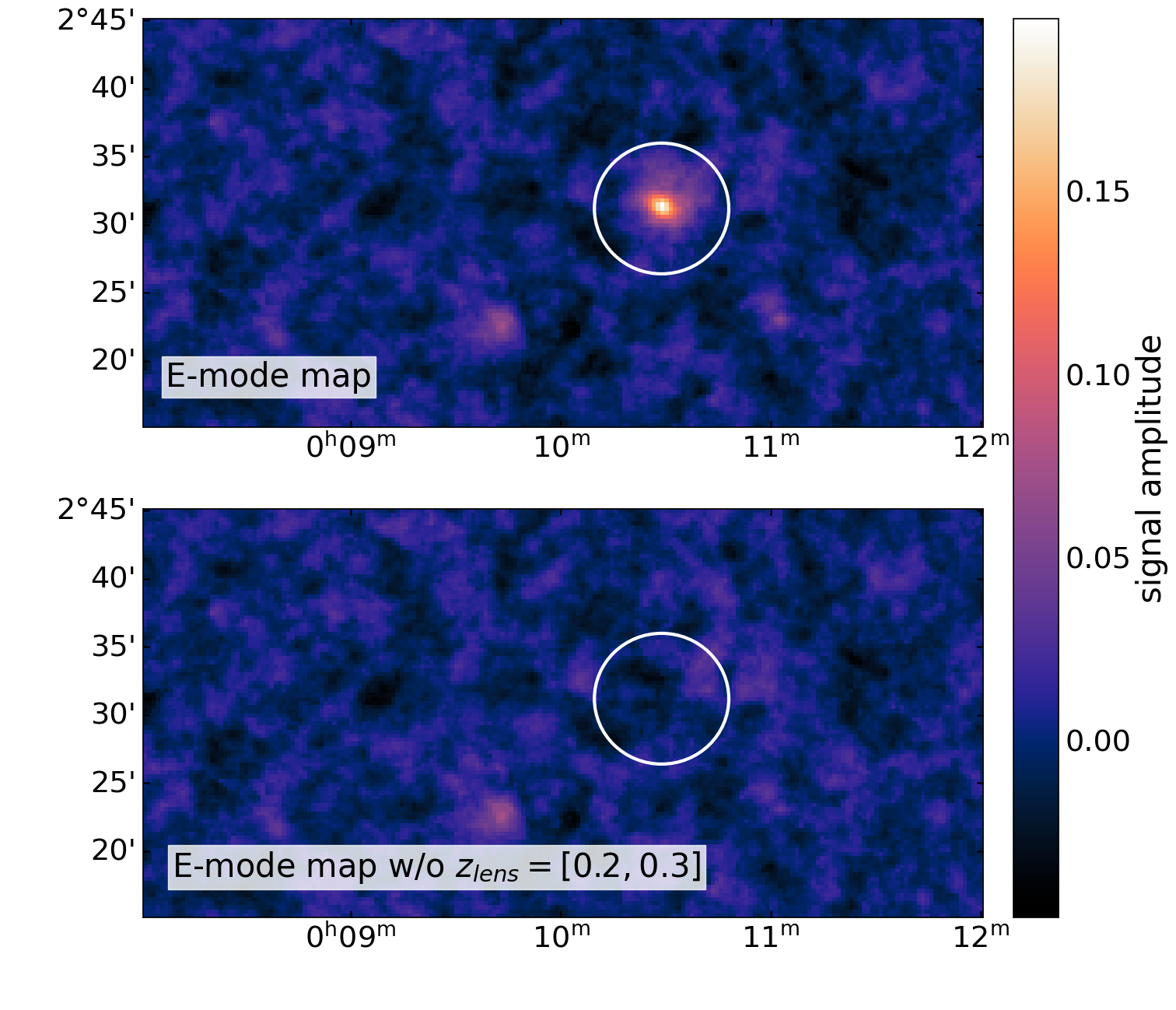}
      \caption{$E$-mode lensing signal map for a 0.5 deg$^2$ region, extracted from the COMPLETE run. 
      Top panel: $E$-mode map including all lens planes. 
      Bottom panel: the same but created by excluding the $z_l \in [0.2, 0.3]$ lens plane, which contains the cluster responsible for the highest peak, whose position is identified by the white circle in both panels.}
         \label{fig:LensPlaneRemoval}
\end{figure}

In this work, we use a different criterion based on a two-step procedure.
First, we identify in the original catalogue the detections that disappear from at least one of the 10 realisations where a redshift slice has been removed: we will refer to them as \textit{vanished detections} from now on. 
This is done using an angular match between the original catalogue of detections and the 10 realisations adopting a matching radius of $0.6$ arcmin to find the ones that have no counterparts. 
The radius has been chosen, studying the typical extension of the peaks in the amplitude map, to be large enough to allow possible positional fluctuations of the maxima of the "lensing blob" but small enough to reduce most of the fortuitous matches between close or even blended detections.
As an example, Fig.~\ref{fig:LensPlaneRemoval} shows a 0.5 deg$^2$ zoom-in of the $E$-mode map resulting from the analysis of the simulation containing all lensing planes (top panel) and the one where the $z_l\in[0.2,0.3]$ slice has been excluded (bottom panel). The large peak visible in the top panel, highlighted by the white circle, disappears completely from the map in the bottom panel, where it is therefore not detected by \texttt{AMICO-WL}. In contrast, all other structures, due to noise and LSS, are basically unaffected.
With this first step, we identified the peaks whose lensing signal is well localised, and thus it is possible to assign a redshift estimate to the vanished detections given by the redshift range of the plane, which, when removed, causes the signal to drop.

The second step consists of matching the vanished detections in each realisation with the haloes present in the corresponding removed redshift slice. By doing so, the matching criterion becomes three-dimensional with a $\Delta z=0.1$ resolution in redshift. For this second step, we choose an angular radius around the haloes of 3 arcmin. The choice of the radius is still based on the typical size expected for the weak lensing imprints in the data but, differently from the first step, we relaxed the size to account for the discretised positions of detections in the amplitude maps and for the presence of a possible shift of the position of the peaks coming from the addition of the weak lensing signal of the LSS present along the line-of-sight.
Let us note that a larger matching radius would create an artificial increase in successful matching due to chance alignments. The matching algorithm also implements an a priori sorting of the halo catalogue by the virial mass and of the detections by $SNR$: high-mass haloes will be first matched with higher $SNR$ peaks inside the matching radius. This step produces the \textit{real detections} catalogue.
 
\begin{figure}
\begin{tabular}{c}
\resizebox{\hsize}{!}{\includegraphics{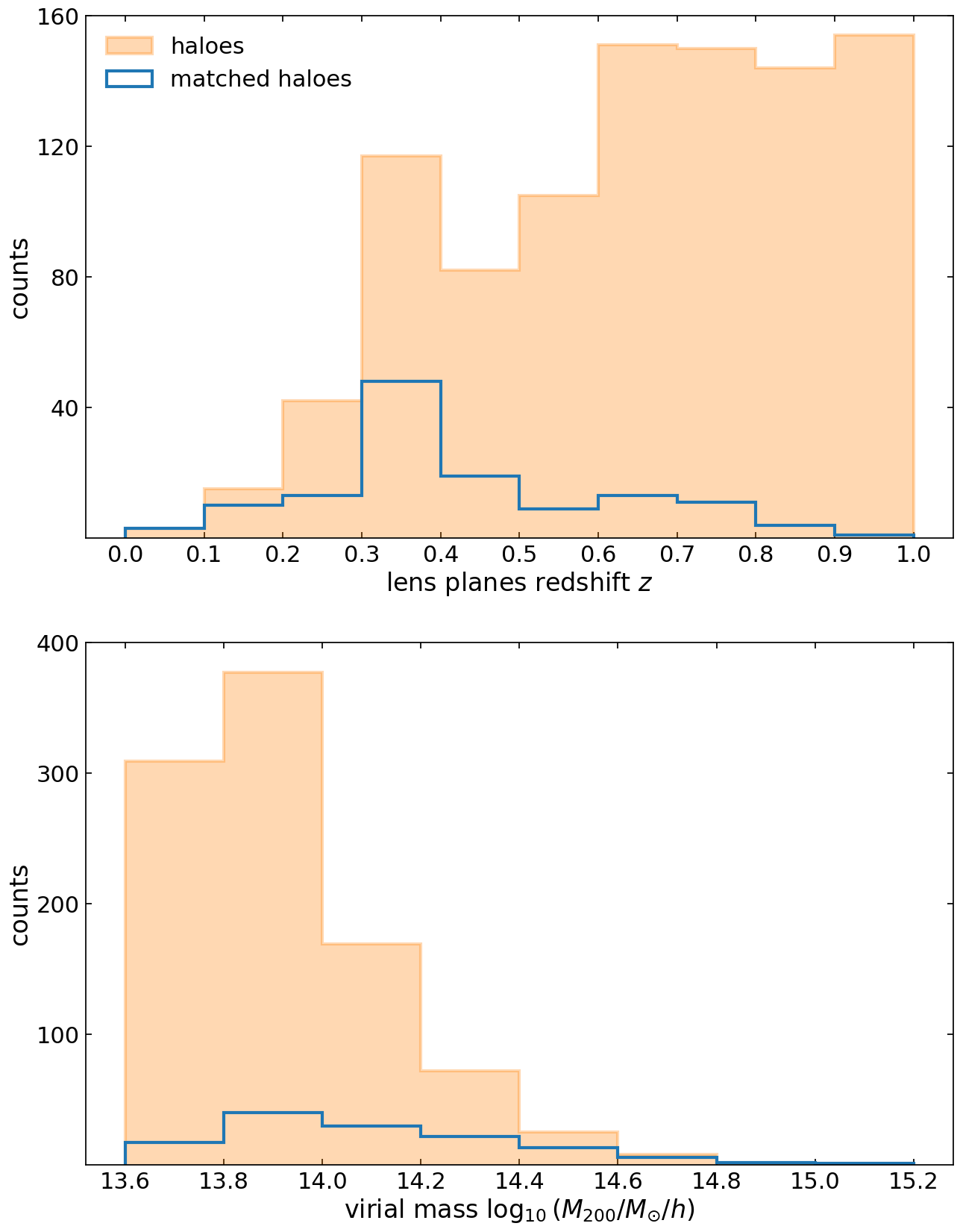}} 
\end{tabular}
\caption{Redshift (top panel) and virial mass (bottom panel) distributions of the simulated matter haloes of the reference catalogue used for the matching procedure (orange histogram). The corresponding distributions of the matched haloes in the COMPLETE run are shown as blue solid line.
}
\label{fig:dist_haloes}
\end{figure}

When matching is performed, we consider haloes with a virial mass $M_{200}>5\times10^{13}
~ h^{-1}\,M_{\odot}$
and redshift $z<1.0$, as it is widely accepted that smaller and/or very distant haloes cannot be detected solely with weak gravitational lensing \citep[see e.g][]{maturi2005optimal, pace2007testing, AndreonBerge2012rich}.
Inside the 5 deg $\times$ 5 deg field of view,
the selected catalogues for the 10 different lens slices, from $z_l\in[0.0,0.1]$ to $z_l\in[0.9,1.0]$ contain 3, 15, 42, 117, 82, 105, 151, 150, 144, 154 dark matter haloes, respectively. This redshift distribution is also shown as a filled orange histogram in the top panel of Fig.~\ref{fig:dist_haloes}, while the corresponding virial mass distribution is displayed in the lower panel.

Note that, in some cases, we associate the same detection to more than one cluster when its $SNR$ falls below the detection threshold for more than one removed lens slice. This can happen when (1) multiple haloes along the line-of-sight contribute cumulatively to a single lensing peak. Removing separately the planes that host these haloes reduces each time the peak signal to below the detection threshold, leading to multiple associations. Alternatively, (2) when the $SNR$ of the peak is close to the threshold, then even a possible small variation in the signal given by noise fluctuations or by the removal of large-scale structures along the line-of-sight can bring the peak below the detection threshold several times so that it is associated with multiple lens planes.
To define unique associations, we associate each detection with the halo with the highest virial mass within the multiple associations to obtain the final catalogue of detected clusters. In the COMPLETE case, less than 7\% of the detections have multiple associations. 

\begin{figure}
\resizebox{\hsize}{!}{\includegraphics{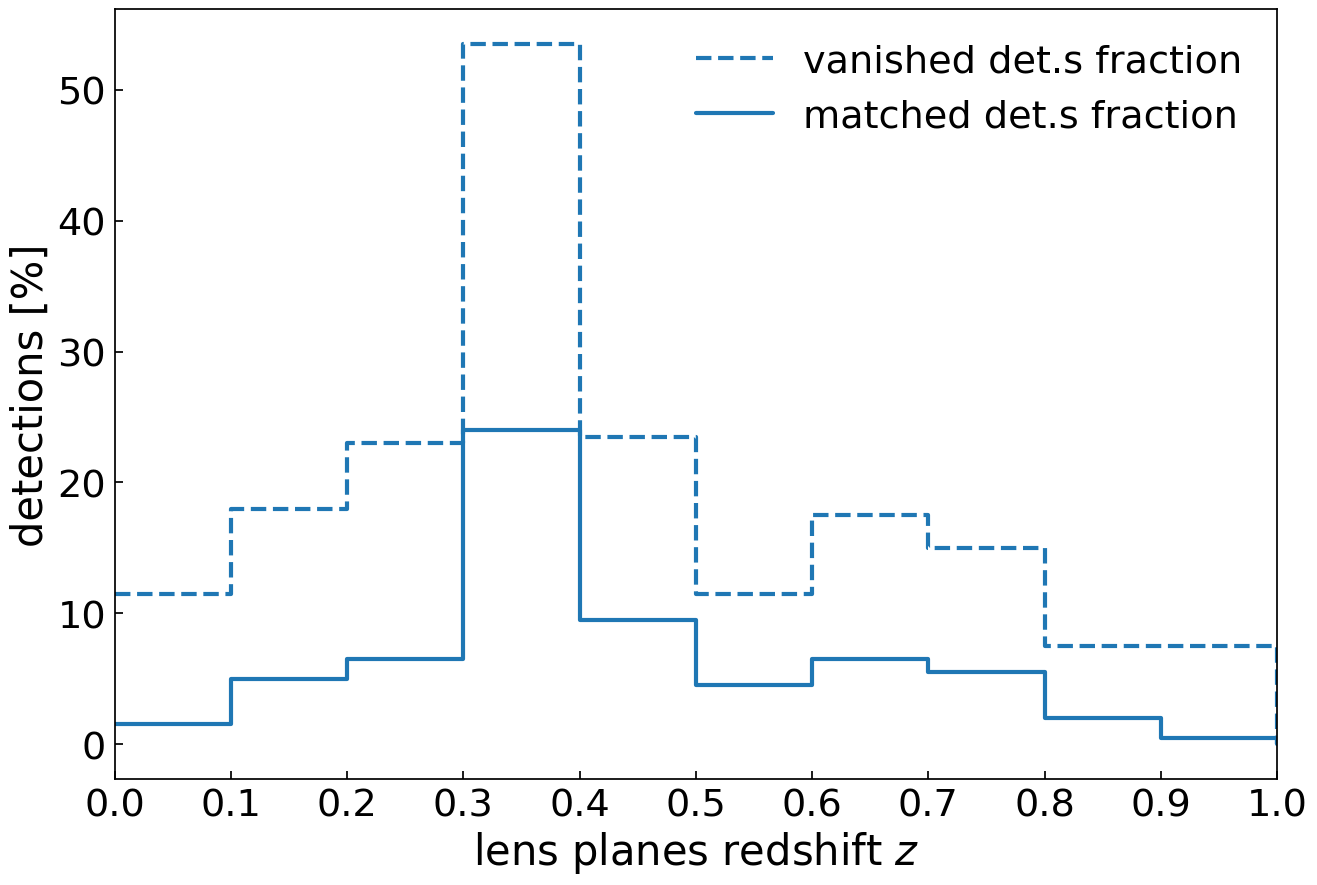}} \\
\caption{Fraction (in percentage) of detections in the primary catalogue which are vanished  (dashed line) and matched (solid line) as a function of the removed lens-plane. The results refer to the COMPLETE run.}
\label{fig:dets_frac}
\end{figure}

Figure~\ref{fig:dets_frac} illustrates how the `blinking' procedure confirms the expectations of lensing theory. Specifically, we show, as a function of the lens plane redshift and for the COMPLETE run only, the fraction of primary detections which are vanished or matched (dashed and solid lines, respectively). These fractions are not constant but peak at $z_l \in [0.3, 0.4]$, consistent with lensing theory, which predicts a stronger lensing signal (and thus a higher detection rate) when the lens is located at an intermediate position between the observer and the sources.

\begin{figure}
\resizebox{\hsize}{!}{\includegraphics{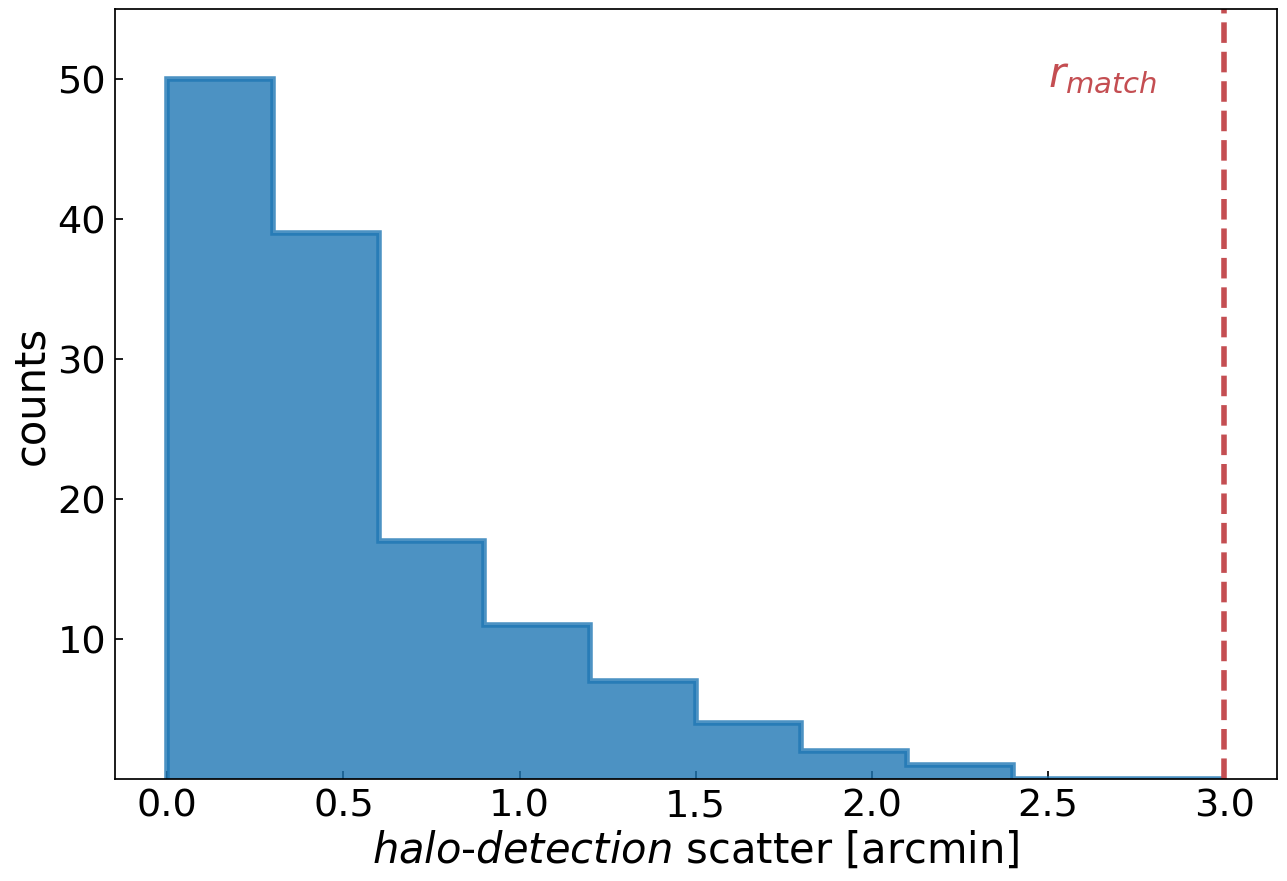}} \\
\caption{Distribution of the scatter in position between the halo and the matched detection in the real detections catalogue. The red dashed vertical line represents the size of the matching radius of 3 arcmin used in the second step of the lens plane matching procedure.
}
\label{fig:scatter}
\end{figure}

In Fig.~\ref{fig:scatter}, we show the distribution of the scatter in position between the detections and the corresponding matched haloes. Even if the matching radius is set to 3 arcmin, the distribution drops well below this value, demonstrating that the number of random matches is negligible, as it would otherwise have produced a flat distribution. This further shows the validity of the adopted matching approach.

The 963 simulated haloes with $M_{200}>5\times10^{13} ~ h^{-1}\,M_{\odot}$
 are matched by the following number of \texttt{AMICO-WL} detections:
\begin{itemize}
    \item \textbf{COMPLETE} run: 131 dark matter haloes, 9 of which have a double or triple association;
    \item \textbf{FR02} run: 132 dark matter haloes, 7 of which have a double association;
    \item \textbf{FR04} run: 141 dark matter haloes, 3 of which have a double association;
    \item \textbf{FR06} run: 129 dark matter haloes, 6 of which have a double association.
\end{itemize}
Detections matching two or three haloes along the line-of-sight usually occur for $SNR\lesssim4.0$, falling into the category of peaks near the detection threshold. The distribution of the matched haloes with respect to the total catalogue is reported in Fig.~\ref{fig:dist_haloes} as a function of redshift (top panel) and virial mass (bottom panel) for the COMPLETE and FR02 runs. Note that the bulk of the dark matter haloes distribution is at low mass and high redshift values, whereas weak lensing detection has a low efficiency. This limitation emphasises the need for complementary data to enhance the sensitivity of weak lensing detections, especially in challenging low-mass, high-redshift environments.


\subsection{Purity and Completeness}

In this section, we describe the results of the statistical analysis performed using the lens-plane-removed matching procedure to the applications of the four filters in \texttt{AMICO-WL} discussed in Sect.~\ref{LPmatch}. In Table~\ref{tab:PUR_COMP}, we report, for the four total samples we produced, the values of the sample completeness, defined as
\begin{equation}
    \text{COMP} = \frac{\text{\# matched haloes}}{\text{\# selected reference haloes}} \;,
\end{equation}
and the purity, defined as
\begin{equation}
    \text{PUR} = \frac{\text{\# real detections}}{\text{\# primary detections}} \;.
\end{equation}
The FR04 and especially the FR02 samples display a higher purity with respect to the COMPLETE one, demonstrating the efficiency of the simple foreground removal we implemented. For the FR02 sample, we find that the purity aligns with the estimates provided in Sect.~\ref{SNR} for the SNR threshold, supporting the validity of the selection procedure.

We also list the $SNR$ thresholds leading to a sample purity of 70\% and 80\% with the corresponding number of detections and completeness. Such a comparison at a fixed sample purity allows for a more coherent comparison of the sample completeness achieved with the different applications. Considering a 70\% purity, the FR02 run delivers the highest completeness (13\% against 6.5\% of the COMPLETE run),  while at 80\% purity, the performance of the four applications is almost equivalent. This is because the selected peaks are those with higher $SNR$ and are related to the haloes with higher virial mass and redshift in the high lensing efficiency range. 

\begin{figure}
\resizebox{\hsize}{!}{\includegraphics{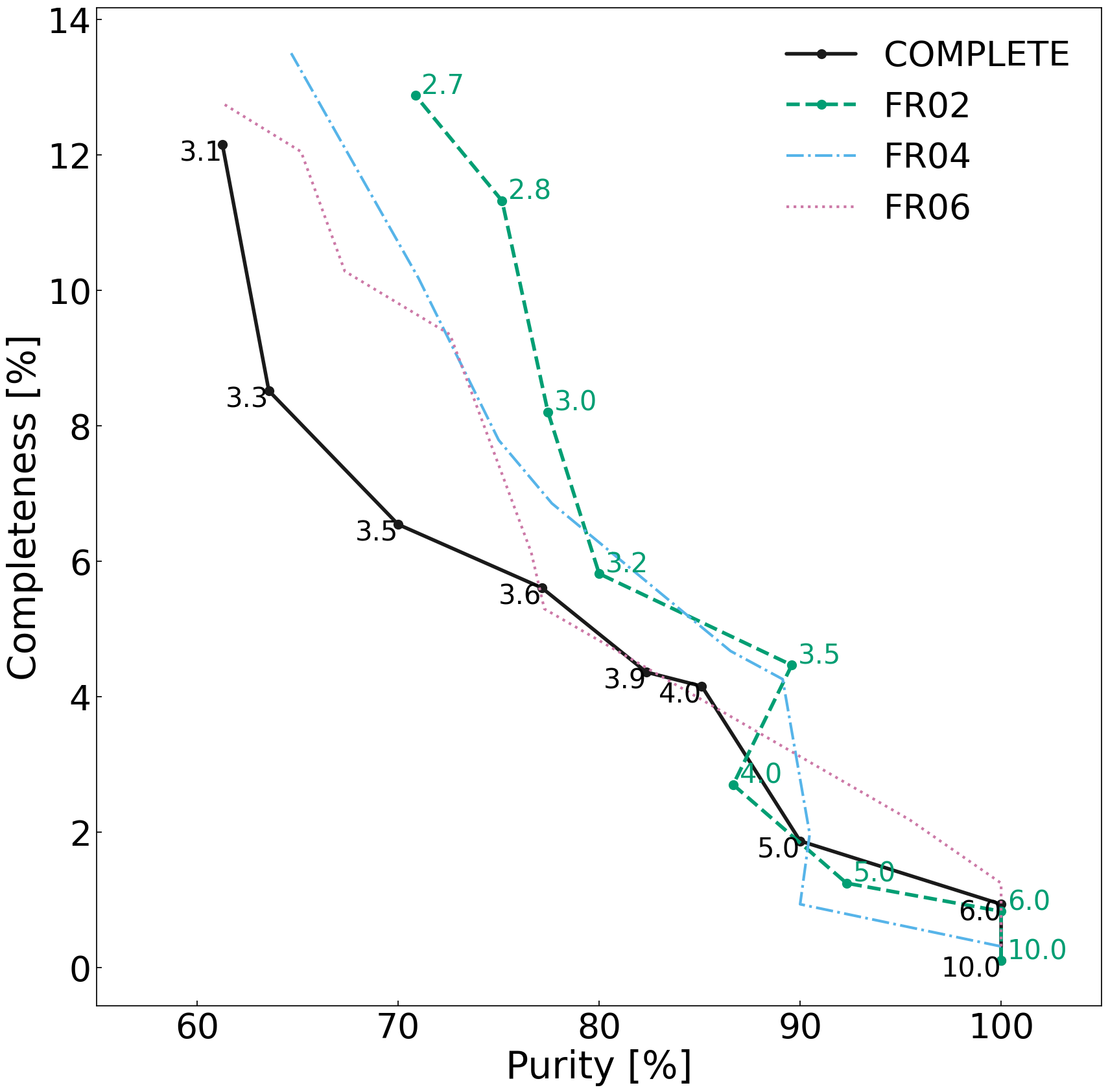}} \\
\caption{Completeness as a function of purity for different $SNR$ thresholds. Different lines refer to the COMPLETE run (black solid line), FR02 run (green dashed line), FR04 run (cyan dash-dotted line) and FR06 run (pink dotted line).  In the case of COMPLETE and FR02, the $SNR$ threshold at which the purity and completeness are computed is noted for each point.}
\label{fig:pur_v_comp}
\end{figure}

In Fig.~\ref{fig:pur_v_comp} we summarise with a ROC curve the performance of \texttt{AMICO-WL} in terms of purity and completeness as a function of $SNR$. As expected, at low $SNR$ more objects are found (high completeness) but with a high probability of detecting spurious sources (low purity). Vice versa, at high $SNR$, only a few (low completeness) but very reliable detections are found (high purity). The runs with foreground removal have a higher completeness than the COMPLETE run for the same purity level. The best performances are delivered by FR02.

\begin{figure*}[t!] 
\begin{subfigure}{0.5\textwidth}
\includegraphics[width=\linewidth, height=0.9\linewidth]{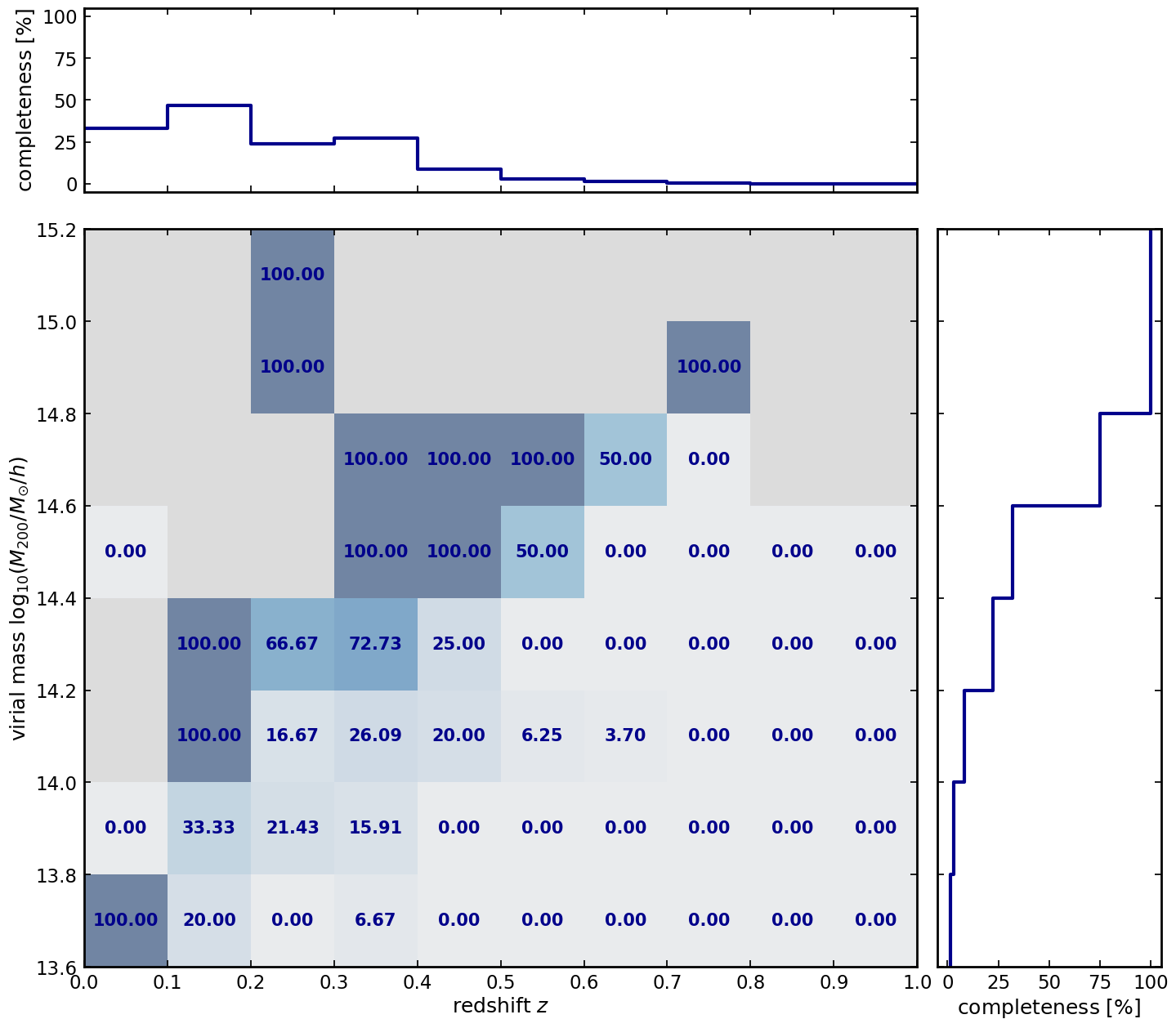}
\end{subfigure}\hspace*{\fill}
\begin{subfigure}{0.5\textwidth}
\includegraphics[width=\linewidth, height=0.9\linewidth]{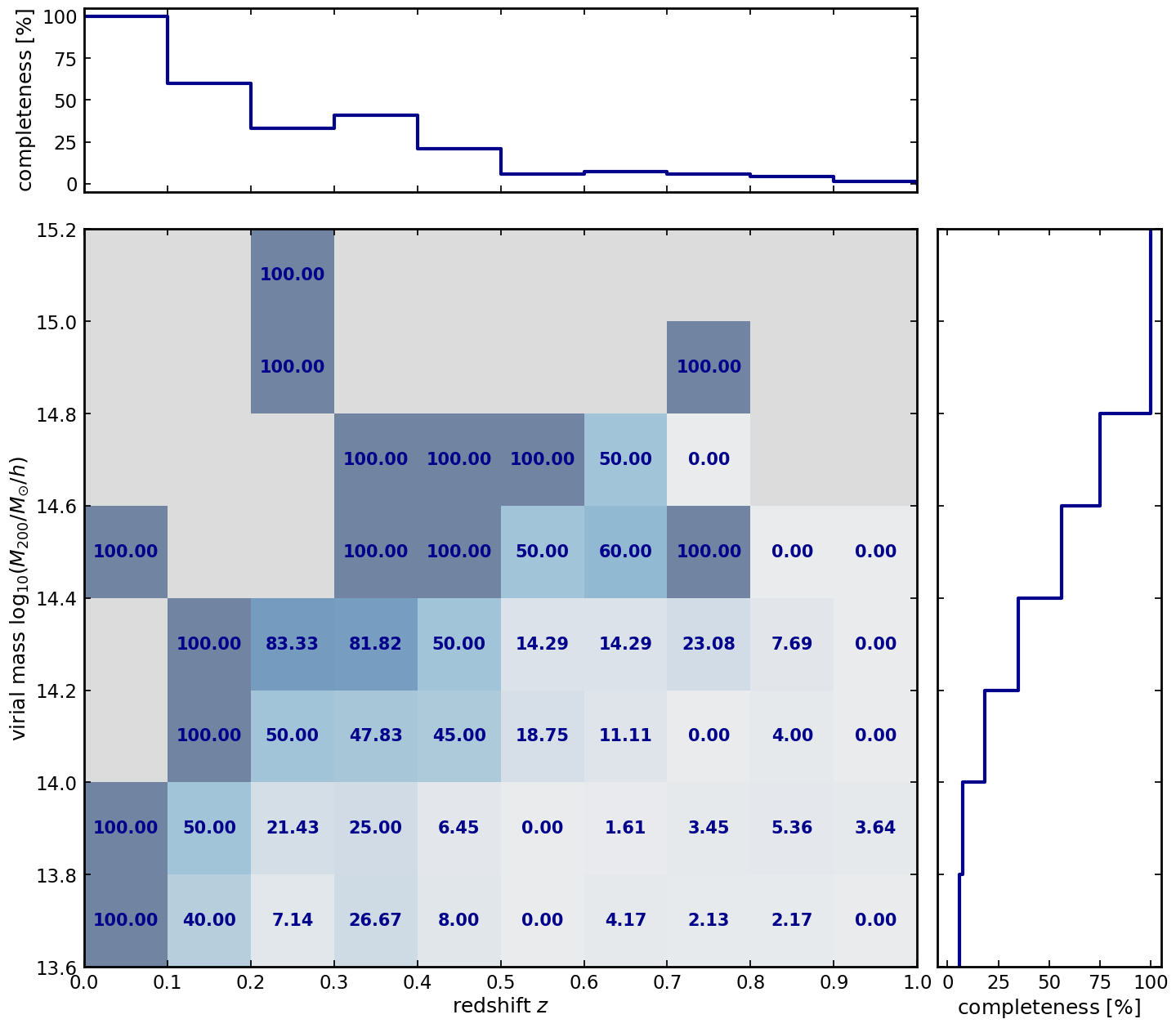}
\end{subfigure}
\caption{The sample completeness in 
two-dimensional bins of redshift and virial mass of the matched haloes for the COMPLETE (left panel) and FR02 (right panel) runs. Both samples are cut at the $SNR$ threshold corresponding to a 70\% purity. The colour map and the values reported inside the two-dimensional bins both refer to completeness. The side plots show the completeness as a function of redshift (on the top) and mass (on the right).} 
\label{fig:comp2D}
\end{figure*}

The completeness of the sample is not uniquely defined, as it depends on the mass cut-off used to define the reference true cluster sample, which can be chosen arbitrarily. It is, therefore, of great interest to investigate the completeness as a function of both the true redshift and the virial mass $M_{200}$ of the matched clusters. This is shown in Fig.~\ref{fig:comp2D} for the COMPLETE and FR02 samples considering a 70\% purity level. In the same figure, the side plots show the completeness as a function of redshift (top panels) and mass (right panels). 
In both cases, the completeness is maximum, 
that is 100\%, in the top left part of the plots that refer to the haloes with high mass ($M_{200}>10^{14.4} 
~ h^{-1}\,M_{\odot}$) and redshift in the range $z=[0.2,0.4]$. This is expected because these are the haloes with the stronger lensing effect: large mass and redshift roughly in between the background source and the observer. The completeness of the FR02 run is higher in all bins, especially for those at lower redshift. 

\section{Analysis of the spurious detections}\label{spurious}

In this section, we investigate the nature of spurious detections to gain an in-depth understanding of the cause of their origin and, at the same time, to check for possible failures in the detection method or in the matching procedure. Here we focus on the results of the COMPLETE run to assess the spurious detections within the standard \texttt{AMICO-WL} run and to further check the efficiency of the foreground removal in avoiding them.

A spurious detection is a galaxy cluster candidate identified by \texttt{AMICO-WL} that does not find a counterpart in the mock catalogue of dark matter haloes through the lens-plane matching procedure. Here we study whether they are noise peaks mistakenly identified as real or false negatives due to failures in the matching procedure. 

\begin{figure}
\resizebox{\hsize}{!}{\includegraphics{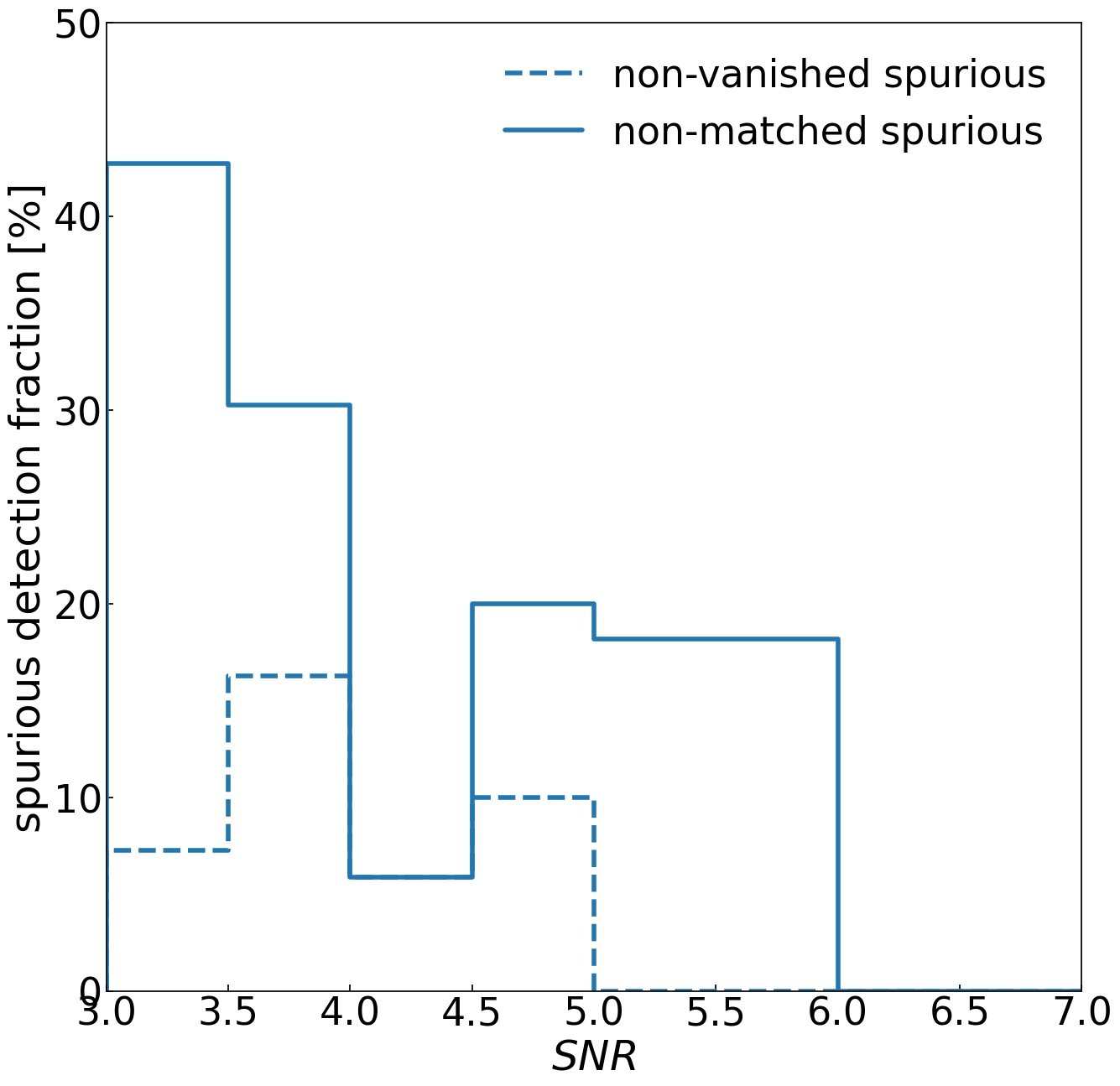}} \\
\caption{Fraction of spurious detections as a function of $SNR$, in the COMPLETE run. The blue dashed line refers to the fraction of the non-vanished spurious, while the blue solid one refers to the total spurious (non-vanished plus non-matched).}
\label{fig:spurious_frac}
\end{figure}

We identify two main categories of spurious detections: the \textit{non-vanished} ones, which are always found regardless of the removal of any of the 10 redshift slices of matter, and the \textit{non-matched} ones, which vanish when one or more lens planes are removed from the ray-tracing but that do not match any dark matter halo in those redshift slices. Of the 82 total spurious detections in the COMPLETE run, 17 are non-vanished and 65 non-matched. Their distribution with respect to $SNR$ is shown in Fig.~\ref{fig:spurious_frac}. Half of the spurious detections have small signal-to-noise ratios, $SNR<4.0$, but several, especially the non-matched ones, have relatively high $SNR$ up to approximately $SNR\approx 6.0$.

The non-vanished spurious are produced by non-compact physical matter distribution or noise fluctuations covering several redshift slices along the line of sight. The latter is the main cause of their origin, as most of them do not have a corresponding convergence peak found in the noiseless maps, confirming their "non-physical" origin.

This is not the case for the two non-vanished spurious detections with the highest $SNR$ values, $SNR=4.88$ and $SNR=4.31$. Their $SNR$ value drops significantly when removing certain redshift slices but always remains above the detection threshold. We found massive dark matter structures along their line-of-sight in the redshift plane in which the $SNR$ decreases. They are, therefore, caused by projection effects due to interlopers.
To account for similar cases, we can relax the selection criterion of the detections performed in the first step of the matching procedure by labelling as also vanished the detections that have more than a 3$\sigma$ variation of $SNR$ evaluated with a K-sigma clipping excluding outliers within the 10 values of $SNR$. The result is that the vanished detections increase for every lens plane, thus reducing the number of non-vanished spurious. Still, the number of real detections does not increase significantly, and the increase in purity is of $\sim 2\%$. We conclude that the negligible increase in the performance does not justify the introduction of this specialised refinement for the identification of vanished detection.  
On the other hand, when cross-checking these two detections in the FR02 run, which is the most effective foreground removal application, we find that the first detection remains spurious, while the second one has a positive match. In total, for the FR02 run, the number of non-vanished spurious decreases from 17 to 6: this shows that reducing the foreground contribution is also effective in decreasing spurious detections.

To investigate the origin of the non-matched spurious detections, we study the distribution of the total mass of the haloes with $M_{200}>10^{12} ~ h^{-1}\,M_{\odot}$ laying along the line-of-sight and within the redshift slice associated with each spurious detection. We thus find that these detections are generated by "compact chains" of low-mass haloes aligned along the line-of-sight with a significant total mass. This analysis revealed additional types of spurious detections, primarily those associated with the redshift plane $z_l\in[0.0,0.1]$.
This redshift plane is sparsely populated by isolated haloes, primarily of mass which is lower than those in the selected reference catalogue used for matching. Despite their low mass, these haloes can still generate $SNR$ peaks comparable to those found in higher-redshift planes. The line-of-sight analysis was also complemented with the addition of the \textit{sub-haloes}. The sub-haloes are bound structures identified by \texttt{SUBFIND} present within the virialised region of dark matter haloes \citep{springel01b,dolag09,giocoli10a}.
We found that a few significant spurious detections arise from sub-haloes excluded from the reference halo catalogue. Including sub-haloes in the reference catalogue could marginally increase completeness by less than $1\%$. However, this inclusion would also significantly increase the complexity of the halo selection process.

To summarise, we identify four main typologies associated with non-matched spurious detections: ($i$) a substantial group without the presence of other structures along the line-ofsight, where $SNR$ fluctuations appear solely due to the extra contribution of observational noise and LSS; ($ii$) several detections associated with a collection of aligned haloes and/or sub-haloes of smaller masses, which, due to the nature of gravitational lensing, can emulate the effect of a single more massive structure, resulting in false positives; ($iii$) detections vanishing in the closest redshift plane, $z_l\in[0.0,0.1]$, in some case with high $SNR$, which are produced by dark matter haloes with masses $M_{200}<5\times10^{13}\, M_{\odot}/h$; ($iv$) a final group of spurious detections arising exclusively from the most massive sub-haloes, which are excluded from the main halo catalogue used for the matching and thus remain unmatched. The spurious detections in groups ($i$) and ($ii$) are related to the nature of the data and the superposition principle holding for weak lensing. In the FR02 run, the number of  \textit{non-matched} spurious detections decreases from 65 to 42, mainly due to the noise reduction that acts on the group ($i$). In contrast, the spurious groups ($iii$) and ($iv$) are false negatives given by the selection criteria employed for the reference catalogue used for matching.


\section{Conclusions}\label{conclusions}

The advent of upcoming wide-field surveys, such as Euclid, will provide an unprecedented wealth of data for studying galaxy clusters through weak lensing. However, effectively extracting cluster information from these massive datasets requires robust and accurate tools for data analysis.
In this work, we introduced \texttt{AMICO-WL}, an extension of the well-tested \texttt{AMICO} algorithm, that allows the detection of galaxy clusters through the weak-gravitational lensing effect rather than the properties of the cluster's galaxies exploited in the original version of the code. The core of the algorithm is the linear optimal matched filter introduced by \citet{maturi2005optimal}, to which we have added several improvements. 

We evaluated the performance of \texttt{AMICO-WL} on a 25 deg$^2$ field mimicking the expected distribution of galaxies in a Euclid-like survey with the simulated shear measurements produced with DUSTGRAIN-{\em pathfinder}. \texttt{AMICO-WL} was applied to the field in four different configurations (COMPLETE, FR02, FR04, FR06) with optimal filters tailored to different redshift cuts of the input catalogue to reduce foreground contamination. 

We used an accurate matching procedure based on the `blinking' of individual redshift planes of the matter simulation to gain redshift information (otherwise missed in weak-lensing analysis) and analysed the code performance in terms of purity and completeness. We then performed a detailed statistical analysis of the spurious detections that are not associated with real sources to better understand their origin. 

These are the main results:
\begin{itemize}

\item 
\texttt{AMICO-WL} applies the robust and well-tested optimal matched filtering technique, originally developed for photometric galaxy cluster detection. Its efficiency is enhanced through the AMICO's computational infrastructure and parallelisation. \texttt{AMICO-WL} is designed to be simple and user-friendly, making it well-suited for application to large datasets while producing reliable results. Future improvements may include a \textit{dynamical filtering mechanism} to vary the sensitivity of the filter for different cluster features (e.g. mass, redshift).

\item 
The detection process includes a cleaning procedure that removes from the $E$-mode amplitude map the lensing signal associated with each detected candidate cluster. This is achieved by subtracting the filtered model signal of the detection at each galaxy position using the same template adopted for the filter construction. The cleaning ensures that overlapping or nearby peaks do not affect subsequent detections, improving the robustness of the identification process.

\item 
The matching procedure identifies halo-detection associations by analysing their response to the removal of individual redshift slices, thus adding redshift information to the weak lensing matching, which allows a reduction of
spurious associations. For the COMPLETE run, 963 simulated haloes with $M_{200} > 5 \times 10^{13} ~ h^{-1}\,M_{\odot}$ are used as a reference, of which 131 objects are matched. This includes 9 detections associated with multiple haloes located along the line-of-sight. The fraction of matched detections peaks at $z_l \in [0.3, 0.4]$, where the lensing efficiency is maximised. We obtained a purity of 61\% for COMPLETE and 71\%, 64\% and 61\% for FR02, FR04 and FR06, respectively.

\item 
The foreground removal technique, based on redshift cuts (with $z_{\rm min} = 0.2, 0.4, 0.6$), reduces the noise coming from low-redshift galaxies, which act as foreground or contribute with very little lensing signal. This foreground removal significantly enhances detection reliability, improving both purity and completeness. For instance:
\begin{itemize}
    \item Considering a 70\% purity level, completeness increases from 6.5\% for the COMPLETE run to 13.0\% for the FR02 run.
    \item Considering a more conservative 80\% purity level, the completeness values for all runs converge since only high-$SNR$ detections related to the most massive haloes remain in the sample.
\end{itemize}
The FR02 run demonstrates higher efficiency in detecting lower-mass haloes compared to the COMPLETE run at similar purity thresholds.
Furthermore, the spurious detection rate in the FR02 run is reduced by 36\% compared to the COMPLETE run. A \textit{dynamic redshift-dependent foreground removal} strategy could further improve both purity and completeness.

\item 
Spurious detections are categorised into \textit{non-vanished} detections, associated with noise or filamentary structures, and \textit{non-matched} detections, often arising from alignments of low-mass haloes along the line-of-sight. Of the 82 spurious detections in the COMPLETE run:
\begin{itemize}
    \item 17 are non-vanished and persist regardless of lens-plane removal.
    \item 65 are non-matched, largely attributed to alignments of low-mass haloes emulating massive structures.
\end{itemize}
From this analysis of the spurious detections, we quantified the power and limits of both the \texttt{AMICO-WL} detection method and the lens plane matching. We suggested some mitigation strategies to reduce the false negatives: refinement of the matching procedure and of the halo catalogue selection. The implementation of these intricate and specialised mitigation measures for false-negative spurious detections is not justified in the general analysis because of the negligible gain in purity they achieve. 

\end{itemize}

In conclusion, \texttt{AMICO-WL} shows its effectiveness as a weak-lensing cluster detection tool, and the matching procedure that has been put in place has proved crucial in determining the characteristics of the structures that have been found. \texttt{AMICO-WL} is currently being evaluated in comparison with other algorithms in the Euclid Weak Lensing Cluster Challenge, demonstrating its versatility and competitiveness. We plan to improve the code by implementing an adaptive filter that would increase the flexibility of the algorithm and allow the detection of dark matter haloes in a broader range of masses and redshift. The ultimate goal is to build an optimised routine for weak-lensing detection of clusters complementary to 
those developed for photometric detection.
The combination of the parent \texttt{AMICO} photometric algorithm with the new \texttt{AMICO-WL} represents an extremely versatile and powerful tool for detecting clusters that have different characteristics.
The next step will be the application of this code to real observational data coming from existing or upcoming wide-field surveys, such as KiDS, CHFTLens, DES, Euclid, and LSST\footnote{\url{https://www.lsst.org/}}
from the Vera C. Rubin Observatory.

\begin{acknowledgements}
LM and CG acknowledge the financial contribution from the PRIN-MUR 2022 20227RNLY3 grant 'The concordance cosmological model: stress-tests with galaxy clusters' supported by Next Generation EU and from the grant ASI n. 2024-10-HH.0 “Attività scientifiche per la missione Euclid – fase E”. 
GC thanks the support from INAF theory Grant 2022: Illuminating Dark Matter using Weak Lensing by Cluster Satellites.
\end{acknowledgements}
%
%

\bibliography{AMICO_WL}

\begin{thebibliography}{115}
\expandafter\ifx\csname natexlab\endcsname\relax\def\natexlab#1{#1}\fi

\bibitem[{Aghanim {et~al.}(2011)Aghanim, Arnaud, Ashdown, Aumont, Baccigalupi,
  Balbi, Banday, Barreiro, Bartelmann, Bartlett, {et~al.}}]{aghanim2011planck}
Aghanim, N., Arnaud, M., Ashdown, M., {et~al.} 2011, Astronomy \& Astrophysics,
  536, A9

\bibitem[{Aiola {et~al.}(2020)Aiola, Calabrese, Maurin, Naess, Schmitt,
  Abitbol, Addison, Ade, Alonso, Amiri, {et~al.}}]{aiola2020atacama}
Aiola, S., Calabrese, E., Maurin, L., {et~al.} 2020, Journal of Cosmology and
  Astroparticle Physics, 2020, 047

\bibitem[{{Allen} {et~al.}(2011){Allen}, {Evrard}, \& {Mantz}}]{allen11}
{Allen}, S.~W., {Evrard}, A.~E., \& {Mantz}, A.~B. 2011, \araa, 49, 409

\bibitem[{{Andreon} \& {Berg{\'e}}(2012)}]{AndreonBerge2012rich}
{Andreon}, S. \& {Berg{\'e}}, J. 2012, \aap, 547, A117

\bibitem[{Andreon {et~al.}(2009)Andreon, Maughan, Trinchieri, \&
  Kurk}]{andreon2009jkcs}
Andreon, S., Maughan, B., Trinchieri, G., \& Kurk, J. 2009, Astronomy \&
  Astrophysics, 507, 147

\bibitem[{Andreon \& Moretti(2011)}]{andreon2011x}
Andreon, S. \& Moretti, A. 2011, Astronomy \& Astrophysics, 536, A37

\bibitem[{Andreon {et~al.}(2024)Andreon, Trinchieri, \&
  Moretti}]{andreon2024observed}
Andreon, S., Trinchieri, G., \& Moretti, A. 2024, Astronomy \& Astrophysics,
  686, A284

\bibitem[{Bartelmann \& Schneider(2001)}]{bartelmann2001weak}
Bartelmann, M. \& Schneider, P. 2001, Physics Reports, 340, 291

\bibitem[{Bellagamba {et~al.}(2011)Bellagamba, Maturi, Hamana, Meneghetti,
  Miyazaki, \& Moscardini}]{bellagamba2011optimal}
Bellagamba, F., Maturi, M., Hamana, T., {et~al.} 2011, Monthly Notices of the
  Royal Astronomical Society, 413, 1145

\bibitem[{Bellagamba {et~al.}(2018)Bellagamba, Roncarelli, Maturi, \&
  Moscardini}]{bellagamba2018amico}
Bellagamba, F., Roncarelli, M., Maturi, M., \& Moscardini, L. 2018, Monthly
  Notices of the Royal Astronomical Society, 473, 5221

\bibitem[{Bertin \& Arnouts(1996)}]{bertin1996sextractor}
Bertin, E. \& Arnouts, S. 1996, Astronomy and astrophysics supplement series,
  117, 393

\bibitem[{Bleem {et~al.}(2020)Bleem, Bocquet, Stalder, Gladders, Ade, Allen,
  Anderson, Annis, Ashby, Austermann, {et~al.}}]{bleem2020sptpol}
Bleem, L., Bocquet, S., Stalder, B., {et~al.} 2020, The Astrophysical Journal
  Supplement Series, 247, 25

\bibitem[{Bleem {et~al.}(2015)Bleem, Stalder, De~Haan, Aird, Allen, Applegate,
  Ashby, Bautz, Bayliss, Benson, {et~al.}}]{bleem2015galaxy}
Bleem, L., Stalder, B., De~Haan, T., {et~al.} 2015, The Astrophysical Journal
  Supplement Series, 216, 27

\bibitem[{B{\"o}hringer {et~al.}(2001)B{\"o}hringer, Schuecker, Guzzo, Collins,
  Voges, Schindler, Neumann, Cruddace, De~Grandi, Chincarini,
  {et~al.}}]{bohringer2001rosat}
B{\"o}hringer, H., Schuecker, P., Guzzo, L., {et~al.} 2001, Astronomy \&
  Astrophysics, 369, 826

\bibitem[{{Boldrin} {et~al.}(2012){Boldrin}, {Giocoli}, {Meneghetti}, \&
  {Moscardini}}]{boldrin12}
{Boldrin}, M., {Giocoli}, C., {Meneghetti}, M., \& {Moscardini}, L. 2012,
  \mnras, 427, 3134

\bibitem[{{Boldrin} {et~al.}(2016){Boldrin}, {Giocoli}, {Meneghetti},
  {Moscardini}, {Tormen}, \& {Biviano}}]{boldrin16}
{Boldrin}, M., {Giocoli}, C., {Meneghetti}, M., {et~al.} 2016, \mnras, 457,
  2738

\bibitem[{Castignani {et~al.}(2022)Castignani, Radovich, Combes, Salom{\'e},
  Maturi, Moscardini, Bardelli, Giocoli, Lesci, Marulli,
  {et~al.}}]{castignani2022star}
Castignani, G., Radovich, M., Combes, F., {et~al.} 2022, Astronomy \&
  Astrophysics, 667, A52

\bibitem[{Castignani {et~al.}(2023)Castignani, Radovich, Combes, Salom{\'e},
  Moscardini, Bardelli, Giocoli, Lesci, Marulli, Maturi,
  {et~al.}}]{castignani2023star}
Castignani, G., Radovich, M., Combes, F., {et~al.} 2023, Astronomy and
  Astrophysics, 672, A139

\bibitem[{{Castro} {et~al.}(2018){Castro}, {Quartin}, {Giocoli}, {Borgani}, \&
  {Dolag}}]{castro17}
{Castro}, T., {Quartin}, M., {Giocoli}, C., {Borgani}, S., \& {Dolag}, K. 2018,
  \mnras, 478, 1305

\bibitem[{Costanzi {et~al.}(2019)Costanzi, Rozo, Simet, Zhang, Evrard, Mantz,
  Rykoff, Jeltema, Gruen, Allen, {et~al.}}]{costanzi2019methods}
Costanzi, M., Rozo, E., Simet, M., {et~al.} 2019, Monthly Notices of the Royal
  Astronomical Society, 488, 4779

\bibitem[{Davis {et~al.}(1985)Davis, Efstathiou, Frenk, \&
  White}]{davis1985evolution}
Davis, M., Efstathiou, G., Frenk, C.~S., \& White, S.~D. 1985, Astrophysical
  Journal, Part 1 (ISSN 0004-637X), vol. 292, May 15, 1985, p. 371-394.
  Research supported by the Science and Engineering Research Council of England
  and NASA., 292, 371

\bibitem[{De~Jong {et~al.}(2017)De~Jong, Kleijn, Erben, Hildebrandt, Kuijken,
  Sikkema, Brescia, Bilicki, Napolitano, Amaro, {et~al.}}]{de2017third}
De~Jong, J.~T., Kleijn, G. A.~V., Erben, T., {et~al.} 2017, Astronomy \&
  Astrophysics, 604, A134

\bibitem[{Dietrich \& Hartlap(2010)}]{dietrich2010cosmology}
Dietrich, J. \& Hartlap, J. 2010, Monthly Notices of the Royal Astronomical
  Society, 402, 1049

\bibitem[{{Dolag} {et~al.}(2009){Dolag}, {Borgani}, {Murante}, \&
  {Springel}}]{dolag09}
{Dolag}, K., {Borgani}, S., {Murante}, G., \& {Springel}, V. 2009, \mnras, 399,
  497

\bibitem[{Dressler(1980)}]{dressler1980galaxy}
Dressler, A. 1980, Astrophysical Journal, Part 1, vol. 236, Mar. 1, 1980, p.
  351-365., 236, 351

\bibitem[{Eckert {et~al.}(2011)Eckert, Molendi, \& Paltani}]{eckert2011cool}
Eckert, D., Molendi, S., \& Paltani, S. 2011, Astronomy \& Astrophysics, 526,
  A79

\bibitem[{Eisenhardt {et~al.}(2008)Eisenhardt, Brodwin, Gonzalez, Stanford,
  Stern, Barmby, Brown, Dawson, Dey, Doi, {et~al.}}]{eisenhardt2008clusters}
Eisenhardt, P.~R., Brodwin, M., Gonzalez, A.~H., {et~al.} 2008, The
  Astrophysical Journal, 684, 905

\bibitem[{{Euclid Collaboration: Adam} {et~al.}(2019){Euclid Collaboration:
  Adam}, R., {Vannier}, {Maurogordato}, {Biviano}, {Adami}, {Ascaso},
  {Bellagamba}, {Benoist}, \& {et al.}}]{adam19}
{Euclid Collaboration: Adam}, R., {Vannier}, M., {et~al.} 2019, \aap, 627, A23

\bibitem[{{Euclid Collaboration: Giocoli} {et~al.}(2024){Euclid Collaboration:
  Giocoli}, C., {Meneghetti}, {Rasia}, {Borgani}, {Despali}, {Lesci},
  {Marulli}, {Moscardini}, {Sereno}, {Cui}, {Knebe}, {Yepes}, {Castro},
  {Corasaniti}, {Pires}, {Castignani}, {Schrabback}, {Pratt}, {Le Brun},
  {Aghanim}, {Amendola}, {Auricchio}, {Baldi}, {Bodendorf}, {Bonino},
  {Branchini}, {Brescia}, {Brinchmann}, {Camera}, {Capobianco}, {Carbone},
  {Carretero}, {Castander}, {Castellano}, {Cavuoti}, {Cledassou}, {Congedo},
  {Conselice}, {Conversi}, {Copin}, {Corcione}, {Courbin}, {Cropper}, {Da
  Silva}, {Degaudenzi}, {Dinis}, {Dubath}, {Dupac}, {Dusini}, {Farrens},
  {Ferriol}, {Fosalba}, {Frailis}, {Franceschi}, {Fumana}, {Galeotta},
  {Garilli}, {Gillis}, {Grazian}, {Grupp}, {Haugan}, {Holmes}, {Hornstrup},
  {Jahnke}, {K{\"u}mmel}, {Kermiche}, {Kilbinger}, {Kunz}, {Kurki-Suonio},
  {Ligori}, {Lilje}, {Lloro}, {Maiorano}, {Mansutti}, {Marggraf}, {Markovic},
  {Massey}, {Maurogordato}, {Mei}, {Merlin}, {Meylan}, {Moresco}, {Munari},
  {Niemi}, {Nightingale}, {Nutma}, {Padilla}, {Paltani}, {Pasian}, {Pedersen},
  {Pettorino}, {Polenta}, {Poncet}, {Popa}, {Raison}, {Renzi}, {Rhodes},
  {Riccio}, {Romelli}, {Roncarelli}, {Rossetti}, {Saglia}, {Sapone},
  {Sartoris}, {Schneider}, {Secroun}, {Serrano}, {Sirignano}, {Sirri},
  {Stanco}, {Starck}, {Tallada-Cresp{\'\i}}, {Taylor}, {Tereno},
  {Toledo-Moreo}, {Torradeflot}, {Tutusaus}, {Valentijn}, {Valenziano},
  {Vassallo}, {Wang}, {Weller}, {Zamorani}, {Zoubian}, {Andreon}, {Bardelli},
  {Boucaud}, {Bozzo}, {Colodro-Conde}, {Di Ferdinando}, {Fabbian}, {Farina},
  {Israel}, {Keih{\"a}nen}, {Lindholm}, {Mauri}, {Neissner}, {Schirmer},
  {Scottez}, {Tenti}, {Zucca}, {Akrami}, {Baccigalupi}, {Ballardini},
  {Bernardeau}, {Biviano}, {Borlaff}, {Burigana}, {Cabanac}, {Cappi},
  {Carvalho}, {Casas}, {Chambers}, {Cooray}, {Courtois}, {Davini}, {de la
  Torre}, {De Lucia}, {Desprez}, {Dole}, {Escartin}, {Escoffier}, {Ferrero},
  {Finelli}, {Gabarra}, {Ganga}, {Garcia-Bellido}, {George}, {Giacomini},
  {Gozaliasl}, {Hildebrandt}, {Hook}, {Jimenez Mu{\~n}oz}, {Joachimi},
  {Kajava}, {Kansal}, {Kirkpatrick}, {Legrand}, {Loureiro}, {Macias-Perez},
  {Magliocchetti}, {Mainetti}, {Maoli}, {Marcin}, {Martinelli}, {Martinet},
  {Martins}, {Matthew}, {Maurin}, {Metcalf}, {Monaco}, {Morgante}, {Nadathur},
  {Nucita}, {Patrizii}, {Peel}, {Pollack}, {Popa}, {Porciani}, {Potter},
  {P{\"o}ntinen}, {Reimberg}, {S{\'a}nchez}, {Sakr}, {Schneider}, {Sefusatti},
  {Shulevski}, {Spurio Mancini}, {Stadel}, {Steinwagner}, {Valiviita},
  {Veropalumbo}, {Viel}, \& {Zinchenko}}]{giocoli24}
{Euclid Collaboration: Giocoli}, C., {Meneghetti}, M., {et~al.} 2024, \aap,
  681, A67

\bibitem[{{Euclid Collaboration: Scaramella} {et~al.}(2022){Euclid
  Collaboration: Scaramella}, R., {Amiaux}, {Mellier}, {Burigana}, {Carvalho},
  {Cuillandre}, {Da Silva}, {Derosa}, {Dinis}, {Maiorano}, {Maris}, {Tereno},
  {Laureijs}, {Boenke}, {Buenadicha}, {Dupac}, {Gaspar Venancio},
  {G{\'o}mez-{\'A}lvarez}, {Hoar}, {Lorenzo Alvarez}, {Racca},
  {Saavedra-Criado}, {Schwartz}, {Vavrek}, {Schirmer}, {Aussel}, {Azzollini},
  {Cardone}, {Cropper}, {Ealet}, {Garilli}, {Gillard}, {Granett}, {Guzzo},
  {Hoekstra}, {Jahnke}, {Kitching}, {Maciaszek}, {Meneghetti}, {Miller},
  {Nakajima}, {Niemi}, {Pasian}, {Percival}, {Pottinger}, {Sauvage},
  {Scodeggio}, {Wachter}, {Zacchei}, {Aghanim}, {Amara}, {Auphan}, {Auricchio},
  {Awan}, {Balestra}, {Bender}, {Bodendorf}, {Bonino}, {Branchini},
  {Brau-Nogue}, {Brescia}, {Candini}, {Capobianco}, {Carbone}, {Carlberg},
  {Carretero}, {Casas}, {Castander}, {Castellano}, {Cavuoti}, {Cimatti},
  {Cledassou}, {Congedo}, {Conselice}, {Conversi}, {Copin}, {Corcione},
  {Costille}, {Courbin}, {Degaudenzi}, {Douspis}, {Dubath}, {Duncan}, {Dusini},
  {Farrens}, {Ferriol}, {Fosalba}, {Fourmanoit}, {Frailis}, {Franceschi},
  {Franzetti}, {Fumana}, {Gillis}, {Giocoli}, {Grazian}, {Grupp}, {Haugan},
  {Holmes}, {Hormuth}, {Hudelot}, {Kermiche}, {Kiessling}, {Kilbinger},
  {Kohley}, {Kubik}, {K{\"u}mmel}, {Kunz}, {Kurki-Suonio}, {Lahav}, {Ligori},
  {Lilje}, {Lloro}, {Mansutti}, {Marggraf}, {Markovic}, {Marulli}, {Massey},
  {Maurogordato}, {Melchior}, {Merlin}, {Meylan}, {Mohr}, {Moresco}, {Morin},
  {Moscardini}, {Munari}, {Nichol}, {Padilla}, {Paltani}, {Peacock},
  {Pedersen}, {Pettorino}, {Pires}, {Poncet}, {Popa}, {Pozzetti}, {Raison},
  {Rebolo}, {Rhodes}, {Rix}, {Roncarelli}, {Rossetti}, {Saglia}, {Schneider},
  {Schrabback}, {Secroun}, {Seidel}, {Serrano}, {Sirignano}, {Sirri},
  {Skottfelt}, {Stanco}, {Starck}, {Tallada-Cresp{\'\i}}, {Tavagnacco},
  {Taylor}, {Teplitz}, {Toledo-Moreo}, {Torradeflot}, {Trifoglio}, {Valentijn},
  {Valenziano}, {Verdoes Kleijn}, {Wang}, {Welikala}, {Weller}, {Wetzstein},
  {Zamorani}, {Zoubian}, {Andreon}, {Baldi}, {Bardelli}, {Boucaud}, {Camera},
  {Di Ferdinando}, {Fabbian}, {Farinelli}, {Galeotta}, {Graci{\'a}-Carpio},
  {Maino}, {Medinaceli}, {Mei}, {Neissner}, {Polenta}, {Renzi}, {Romelli},
  {Rosset}, {Sureau}, {Tenti}, {Vassallo}, {Zucca}, {Baccigalupi},
  {Balaguera-Antol{\'\i}nez}, {Battaglia}, {Biviano}, {Borgani}, {Bozzo},
  {Cabanac}, {Cappi}, {Casas}, {Castignani}, {Colodro-Conde}, {Coupon},
  {Courtois}, {Cuby}, {de la Torre}, {Desai}, {Dole}, {Fabricius}, {Farina},
  {Ferreira}, {Finelli}, {Flose-Reimberg}, {Fotopoulou}, {Ganga}, {Gozaliasl},
  {Hook}, {Keihanen}, {Kirkpatrick}, {Liebing}, {Lindholm}, {Mainetti},
  {Martinelli}, {Martinet}, {Maturi}, {McCracken}, {Metcalf}, {Morgante},
  {Nightingale}, {Nucita}, {Patrizii}, {Potter}, {Riccio}, {S{\'a}nchez},
  {Sapone}, {Schewtschenko}, {Schultheis}, {Scottez}, {Teyssier}, {Tutusaus},
  {Valiviita}, {Viel}, {Vriend}, \& {Whittaker}}]{scaramella22}
{Euclid Collaboration: Scaramella}, R., {Amiaux}, J., {et~al.} 2022, \aap, 662,
  A112

\bibitem[{Fan {et~al.}(2010)Fan, Shan, \& Liu}]{fan2010noisy}
Fan, Z., Shan, H., \& Liu, J. 2010, The Astrophysical Journal, 719, 1408

\bibitem[{Farrens(2011)}]{farrens2011optical}
Farrens, S. 2011, PhD thesis, UCL (University College London)

\bibitem[{Gavazzi \& Soucail(2007)}]{gavazzi2007weak}
Gavazzi, R. \& Soucail, G. 2007, Astronomy \& Astrophysics, 462, 459

\bibitem[{George {et~al.}(2011)George, Leauthaud, Bundy, Finoguenov, Tinker,
  Lin, Mei, Kneib, Aussel, Behroozi, {et~al.}}]{george2011galaxies}
George, M.~R., Leauthaud, A., Bundy, K., {et~al.} 2011, The Astrophysical
  Journal, 742, 125

\bibitem[{Ghirardini {et~al.}(2024)Ghirardini, Bulbul, Artis, Clerc, Garrel,
  Grandis, Kluge, Liu, Bahar, Balzer, {et~al.}}]{ghirardini2024srg}
Ghirardini, V., Bulbul, E., Artis, E., {et~al.} 2024, arXiv preprint
  arXiv:2402.08458

\bibitem[{Gilbank {et~al.}(2011)Gilbank, Gladders, Yee, \&
  Hsieh}]{gilbank2011red}
Gilbank, D.~G., Gladders, M., Yee, H., \& Hsieh, B. 2011, The Astronomical
  Journal, 141, 94

\bibitem[{{Giocoli} {et~al.}(2018{\natexlab{a}}){Giocoli}, {Baldi}, \&
  {Moscardini}}]{giocoli18b}
{Giocoli}, C., {Baldi}, M., \& {Moscardini}, L. 2018{\natexlab{a}}, \mnras,
  481, 2813

\bibitem[{Giocoli {et~al.}(2021)Giocoli, Marulli, Moscardini, Sereno,
  Veropalumbo, Gigante, Maturi, Radovich, Bellagamba, Roncarelli,
  {et~al.}}]{giocoli2021amico}
Giocoli, C., Marulli, F., Moscardini, L., {et~al.} 2021, Astronomy \&
  Astrophysics, 653, A19

\bibitem[{{Giocoli} {et~al.}(2015){Giocoli}, {Metcalf}, {Baldi}, {Meneghetti},
  {Moscardini}, \& {Petkova}}]{giocoli15}
{Giocoli}, C., {Metcalf}, R.~B., {Baldi}, M., {et~al.} 2015, \mnras, 452, 2757

\bibitem[{{Giocoli} {et~al.}(2018{\natexlab{b}}){Giocoli}, {Moscardini},
  {Baldi}, {Meneghetti}, \& {Metcalf}}]{giocoli18a}
{Giocoli}, C., {Moscardini}, L., {Baldi}, M., {Meneghetti}, M., \& {Metcalf},
  R.~B. 2018{\natexlab{b}}, \mnras [\eprint[arXiv]{1801.01886}]

\bibitem[{{Giocoli} {et~al.}(2010a){Giocoli}, {Tormen}, {Sheth}, \& {van den
  Bosch}}]{giocoli10a}
{Giocoli}, C., {Tormen}, G., {Sheth}, R.~K., \& {van den Bosch}, F.~C. 2010a,
  \mnras, 404, 502

\bibitem[{Hamana {et~al.}(2012)Hamana, Oguri, Shirasaki, \&
  Sato}]{hamana2012scatter}
Hamana, T., Oguri, M., Shirasaki, M., \& Sato, M. 2012, Monthly Notices of the
  Royal Astronomical Society, 425, 2287

\bibitem[{Hamana {et~al.}(2020)Hamana, Shirasaki, \& Lin}]{hamana2020weak}
Hamana, T., Shirasaki, M., \& Lin, Y.-T. 2020, Publications of the Astronomical
  Society of Japan, 72, 78

\bibitem[{Hamana {et~al.}(2004)Hamana, Takada, \&
  Yoshida}]{hamana2004searching}
Hamana, T., Takada, M., \& Yoshida, N. 2004, Monthly Notices of the Royal
  Astronomical Society, 350, 893

\bibitem[{Hasselfield {et~al.}(2013)Hasselfield, Hilton, Marriage, Addison,
  Barrientos, Battaglia, Battistelli, Bond, Crichton, Das,
  {et~al.}}]{hasselfield2013atacama}
Hasselfield, M., Hilton, M., Marriage, T.~A., {et~al.} 2013, Journal of
  Cosmology and Astroparticle Physics, 2013, 008

\bibitem[{Hennawi \& Spergel(2005)}]{hennawi2005shear}
Hennawi, J.~F. \& Spergel, D.~N. 2005, The Astrophysical Journal, 624, 59

\bibitem[{Hetterscheidt {et~al.}(2005)Hetterscheidt, Erben, Schneider, Maoli,
  Van~Waerbeke, \& Mellier}]{hetterscheidt2005searching}
Hetterscheidt, M., Erben, T., Schneider, P., {et~al.} 2005, Astronomy \&
  Astrophysics, 442, 43

\bibitem[{{Hilbert} {et~al.}(2020){Hilbert}, {Barreira}, {Fabbian}, {Fosalba},
  {Giocoli}, {Bose}, {Calabrese}, {Carbone}, {Davies}, {Li}, {Llinares}, \&
  {Monaco}}]{hilbert20}
{Hilbert}, S., {Barreira}, A., {Fabbian}, G., {et~al.} 2020, \mnras, 493, 305

\bibitem[{Ingoglia {et~al.}(2022)Ingoglia, Covone, Sereno, Giocoli, Bardelli,
  Bellagamba, Castignani, Farrens, Hildebrandt, Joudaki,
  {et~al.}}]{ingoglia2022amico}
Ingoglia, L., Covone, G., Sereno, M., {et~al.} 2022, Monthly Notices of the
  Royal Astronomical Society, 511, 1484

\bibitem[{Jarvis {et~al.}(2004)Jarvis, Bernstein, \& Jain}]{jarvis2004skewness}
Jarvis, M., Bernstein, G., \& Jain, B. 2004, Monthly Notices of the Royal
  Astronomical Society, 352, 338

\bibitem[{Kepner {et~al.}(1999)Kepner, Fan, Bahcall, Gunn, Lupton, \&
  Xu}]{kepner1999automated}
Kepner, J., Fan, X., Bahcall, N., {et~al.} 1999, The Astrophysical Journal,
  517, 78

\bibitem[{Koester {et~al.}(2007)Koester, McKay, Annis, Wechsler, Evrard, Bleem,
  Becker, Johnston, Sheldon, Nichol, {et~al.}}]{koester2007maxbcg}
Koester, B.~P., McKay, T.~A., Annis, J., {et~al.} 2007, The Astrophysical
  Journal, 660, 239

\bibitem[{Koulouridis {et~al.}(2021)Koulouridis, Clerc, Sadibekova, Chira,
  Drigga, Faccioli, Le~F{\`e}vre, Garrel, Gaynullina, Gkini,
  {et~al.}}]{koulouridis2021x}
Koulouridis, E., Clerc, N., Sadibekova, T., {et~al.} 2021, Astronomy \&
  Astrophysics, 652, A12

\bibitem[{Kuchner {et~al.}(2017)Kuchner, Ziegler, Verdugo, Bamford, \&
  H{\"a}u{\ss}ler}]{kuchner2017effects}
Kuchner, U., Ziegler, B., Verdugo, M., Bamford, S., \& H{\"a}u{\ss}ler, B.
  2017, Astronomy \& Astrophysics, 604, A54

\bibitem[{Laureijs {et~al.}(2011)Laureijs, Amiaux, Arduini, Augueres,
  Brinchmann, Cole, Cropper, Dabin, Duvet, Ealet,
  {et~al.}}]{laureijs2011euclid}
Laureijs, R., Amiaux, J., Arduini, S., {et~al.} 2011, arXiv preprint
  arXiv:1110.3193

\bibitem[{Leauthaud {et~al.}(2007)Leauthaud, Massey, Kneib, Rhodes, Johnston,
  Capak, Heymans, Ellis, Koekemoer, Le~Fevre, {et~al.}}]{leauthaud2007weak}
Leauthaud, A., Massey, R., Kneib, J.-P., {et~al.} 2007, The Astrophysical
  Journal Supplement Series, 172, 219

\bibitem[{Lesci {et~al.}(2022{\natexlab{a}})Lesci, Marulli, Moscardini, Sereno,
  Veropalumbo, Maturi, Giocoli, Radovich, Bellagamba, Roncarelli,
  {et~al.}}]{lesci2022amicoa}
Lesci, G., Marulli, F., Moscardini, L., {et~al.} 2022{\natexlab{a}}, Astronomy
  \& Astrophysics, 659, A88

\bibitem[{Lesci {et~al.}(2022{\natexlab{b}})Lesci, Nanni, Marulli, Moscardini,
  Veropalumbo, Maturi, Sereno, Radovich, Bellagamba, Roncarelli,
  {et~al.}}]{lesci2022amicob}
Lesci, G., Nanni, L., Marulli, F., {et~al.} 2022{\natexlab{b}}, Astron.
  Astrophys, 665, A100

\bibitem[{Lewis {et~al.}(2000)Lewis, Challinor, \& Lasenby}]{camb}
Lewis, A., Challinor, A., \& Lasenby, A. 2000, Astrophys. J., 538, 473

\bibitem[{Lin {et~al.}(2016)Lin, Kilbinger, \& Pires}]{lin2016new}
Lin, C.-A., Kilbinger, M., \& Pires, S. 2016, Astronomy \& Astrophysics, 593,
  A88

\bibitem[{Marulli {et~al.}(2018)Marulli, Veropalumbo, Sereno, Moscardini,
  Pacaud, Pierre, Plionis, Cappi, Adami, Alis, {et~al.}}]{marulli2018xxl}
Marulli, F., Veropalumbo, A., Sereno, M., {et~al.} 2018, Astronomy \&
  Astrophysics, 620, A1

\bibitem[{Maturi {et~al.}(2019)Maturi, Bellagamba, Radovich, Roncarelli,
  Sereno, Moscardini, Bardelli, \& Puddu}]{maturi2019amico}
Maturi, M., Bellagamba, F., Radovich, M., {et~al.} 2019, Monthly Notices of the
  Royal Astronomical Society, 485, 498

\bibitem[{Maturi {et~al.}(2023)Maturi, Finoguenov, Lopes, Delgado, Dupke,
  Cypriano, Carrasco, Diego, Penna-Lima, Doubrawa,
  {et~al.}}]{maturi2023minijpas}
Maturi, M., Finoguenov, A., Lopes, P., {et~al.} 2023, Astronomy \&
  Astrophysics, 678, A145

\bibitem[{Maturi {et~al.}(2005)Maturi, Meneghetti, Bartelmann, Dolag, \&
  Moscardini}]{maturi2005optimal}
Maturi, M., Meneghetti, M., Bartelmann, M., Dolag, K., \& Moscardini, L. 2005,
  Astronomy \& Astrophysics, 442, 851

\bibitem[{Maturi {et~al.}(2007)Maturi, Schirmer, Meneghetti, Bartelmann, \&
  Moscardini}]{maturi2007searching}
Maturi, M., Schirmer, M., Meneghetti, M., Bartelmann, M., \& Moscardini, L.
  2007, Astronomy \& Astrophysics, 462, 473

\bibitem[{Mellier {et~al.}(2024)Mellier, Barroso, Ach{\'u}carro, Adamek, Adam,
  Addison, Aghanim, Aguena, Ajani, Akrami, {et~al.}}]{mellier2024euclid}
Mellier, Y., Barroso, J., Ach{\'u}carro, A., {et~al.} 2024, arXiv preprint
  arXiv:2405.13491

\bibitem[{Miyazaki {et~al.}(2007)Miyazaki, Hamana, Ellis, Kashikawa, Massey,
  Taylor, \& Refregier}]{miyazaki2007subaru}
Miyazaki, S., Hamana, T., Ellis, R.~S., {et~al.} 2007, The Astrophysical
  Journal, 669, 714

\bibitem[{Miyazaki {et~al.}(2002)Miyazaki, Hamana, Shimasaku, Furusawa, Doi,
  Hamabe, Imi, Kimura, Komiyama, Nakata, {et~al.}}]{miyazaki2002searching}
Miyazaki, S., Hamana, T., Shimasaku, K., {et~al.} 2002, The Astrophysical
  Journal, 580, L97

\bibitem[{Miyazaki {et~al.}(2018)Miyazaki, Oguri, Hamana, Shirasaki, Koike,
  Komiyama, Umetsu, Utsumi, Okabe, More, {et~al.}}]{miyazaki2018large}
Miyazaki, S., Oguri, M., Hamana, T., {et~al.} 2018, Publications of the
  Astronomical Society of Japan, 70, S27

\bibitem[{Navarro(1996)}]{navarro1996structure}
Navarro, J.~F. 1996, in Symposium-international astronomical union, Vol. 171,
  Cambridge University Press, 255--258

\bibitem[{Navarro {et~al.}(2004)Navarro, Hayashi, Power, Jenkins, Frenk, White,
  Springel, Stadel, \& Quinn}]{navarro2004inner}
Navarro, J.~F., Hayashi, E., Power, C., {et~al.} 2004, Monthly Notices of the
  Royal Astronomical Society, 349, 1039

\bibitem[{Oguri {et~al.}(2021)Oguri, Miyazaki, Li, Luo, Mitsuishi, Miyatake,
  More, Nishizawa, Okabe, Ota, {et~al.}}]{oguri2021hundreds}
Oguri, M., Miyazaki, S., Li, X., {et~al.} 2021, Publications of the
  Astronomical Society of Japan, 73, 817

\bibitem[{Pacaud {et~al.}(2016)Pacaud, Clerc, Giles, Adami, Sadibekova, Pierre,
  Maughan, Lieu, Le~F{\`e}vre, Alis, {et~al.}}]{pacaud2016xxl}
Pacaud, F., Clerc, N., Giles, P., {et~al.} 2016, Astronomy \& Astrophysics,
  592, A2

\bibitem[{Pace {et~al.}(2007)Pace, Maturi, Meneghetti, Bartelmann, Moscardini,
  \& Dolag}]{pace2007testing}
Pace, F., Maturi, M., Meneghetti, M., {et~al.} 2007, Astronomy \& Astrophysics,
  471, 731

\bibitem[{Piffaretti {et~al.}(2011)Piffaretti, Arnaud, Pratt, Pointecouteau, \&
  Melin}]{piffaretti2011mcxc}
Piffaretti, R., Arnaud, M., Pratt, G., Pointecouteau, E., \& Melin, J.-B. 2011,
  Astronomy \& Astrophysics, 534, A109

\bibitem[{Pires {et~al.}(2020)Pires, Vandenbussche, Kansal, Bender, Blot,
  Bonino, Boucaud, Brinchmann, Capobianco, Carretero,
  {et~al.}}]{pires2020euclid}
Pires, S., Vandenbussche, V., Kansal, V., {et~al.} 2020, Astronomy \&
  Astrophysics, 638, A141

\bibitem[{{Planck Collaboration} {et~al.}(2015){Planck Collaboration}, {Adam},
  {Ade}, {Aghanim}, {Akrami}, {Alves}, {Arnaud}, {Arroja}, {Aumont},
  {Baccigalupi}, \& et~al.}]{planck1_15}
{Planck Collaboration}, {Adam}, R., {Ade}, P.~A.~R., {et~al.} 2015, ArXiv
  e-prints: 1502.01582 [\eprint[arXiv]{1502.01582}]

\bibitem[{{Planck Collaboration} {et~al.}(2011){Planck Collaboration}, {Ade},
  {Aghanim}, {Arnaud}, {Ashdown}, {Aumont}, {Baccigalupi}, {Baker}, {Balbi},
  {Banday}, \& et~al.}]{planck1_11}
{Planck Collaboration}, {Ade}, P.~A.~R., {Aghanim}, N., {et~al.} 2011, \aap,
  536, A1

\bibitem[{Puddu {et~al.}(2021)Puddu, Radovich, Sereno, Bardelli, Maturi,
  Moscardini, Bellagamba, Giocoli, Marulli, \& Roncarelli}]{puddu2021amico}
Puddu, E., Radovich, M., Sereno, M., {et~al.} 2021, Astronomy \& Astrophysics,
  645, A9

\bibitem[{Pyne~III(1995)}]{pyne1995null}
Pyne~III, E.~W. 1995, Null geodesics in perturbed spacetimes (Harvard
  University)

\bibitem[{Radovich {et~al.}(2020)Radovich, Tortora, Bellagamba, Maturi,
  Moscardini, Puddu, Roncarelli, Roy, Bardelli, Marulli,
  {et~al.}}]{radovich2020amico}
Radovich, M., Tortora, C., Bellagamba, F., {et~al.} 2020, Monthly Notices of
  the Royal Astronomical Society, 498, 4303

\bibitem[{Ramella {et~al.}(2001)Ramella, Boschin, Fadda, \&
  Nonino}]{ramella2001finding}
Ramella, M., Boschin, W., Fadda, D., \& Nonino, M. 2001, Astronomy \&
  Astrophysics, 368, 776

\bibitem[{Rettura {et~al.}(2014)Rettura, Martinez-Manso, Stern, Mei, Ashby,
  Brodwin, Gettings, Gonzalez, Stanford, \& Bartlett}]{rettura2014candidate}
Rettura, A., Martinez-Manso, J., Stern, D., {et~al.} 2014, The Astrophysical
  Journal, 797, 109

\bibitem[{Romanello {et~al.}(2024)Romanello, Marulli, Moscardini, Lesci,
  Sartoris, Contarini, Giocoli, Bardelli, Busillo, Castignani,
  {et~al.}}]{romanello2024amico}
Romanello, M., Marulli, F., Moscardini, L., {et~al.} 2024, Astronomy \&
  Astrophysics, 682, A72

\bibitem[{{Roncarelli} {et~al.}(2007){Roncarelli}, {Moscardini}, {Borgani}, \&
  {Dolag}}]{roncarelli07}
{Roncarelli}, M., {Moscardini}, L., {Borgani}, S., \& {Dolag}, K. 2007, \mnras,
  378, 1259

\bibitem[{Rosati {et~al.}(2002)Rosati, Borgani, \&
  Norman}]{rosati2002evolution}
Rosati, P., Borgani, S., \& Norman, C. 2002, Annual Review of Astronomy and
  Astrophysics, 40, 539

\bibitem[{{Rozo} {et~al.}(2010){Rozo}, {Wechsler}, {Rykoff}, {Annis}, {Becker},
  {Evrard}, {Frieman}, {Hansen}, \& {et~al.}}]{rozo10}
{Rozo}, E., {Wechsler}, R.~H., {Rykoff}, E.~S., {et~al.} 2010, \apj, 708, 645

\bibitem[{Rykoff {et~al.}(2014)Rykoff, Rozo, Busha, Cunha, Finoguenov, Evrard,
  Hao, Koester, Leauthaud, Nord, {et~al.}}]{rykoff2014redmapper}
Rykoff, E., Rozo, E., Busha, M., {et~al.} 2014, The Astrophysical Journal, 785,
  104

\bibitem[{Sadibekova {et~al.}(2024)Sadibekova, Arnaud, Pratt, Tarr{\'\i}o, \&
  Melin}]{sadibekova2024mcxc}
Sadibekova, T., Arnaud, M., Pratt, G., Tarr{\'\i}o, P., \& Melin, J.-B. 2024,
  Astronomy \& Astrophysics, 688, A187

\bibitem[{Sarron {et~al.}(2018)Sarron, Martinet, Durret, \&
  Adami}]{sarron2018evolution}
Sarron, F., Martinet, N., Durret, F., \& Adami, C. 2018, Astronomy \&
  Astrophysics, 613, A67

\bibitem[{Schaffer {et~al.}(2011)Schaffer, Crawford, Aird, Benson, Bleem,
  Carlstrom, Chang, Cho, Crites, De~Haan, {et~al.}}]{schaffer2011first}
Schaffer, K., Crawford, T., Aird, K., {et~al.} 2011, The Astrophysical Journal,
  743, 90

\bibitem[{Schirmer {et~al.}(2007)Schirmer, Erben, Hetterscheidt, \&
  Schneider}]{schirmer2007gabods}
Schirmer, M., Erben, T., Hetterscheidt, M., \& Schneider, P. 2007, Astronomy \&
  Astrophysics, 462, 875

\bibitem[{Schirmer {et~al.}(2004)Schirmer, Erben, Schneider, Wolf, \&
  Meisenheimer}]{schirmer2004gabods}
Schirmer, M., Erben, T., Schneider, P., Wolf, C., \& Meisenheimer, K. 2004,
  Astronomy \& Astrophysics, 420, 75

\bibitem[{Schneider(1996)}]{schneider1996detection}
Schneider, P. 1996, Monthly Notices of the Royal Astronomical Society, 283, 837

\bibitem[{Schneider {et~al.}(1992)Schneider, Ehlers, Falco, Schneider, Ehlers,
  \& Falco}]{schneider1992properties}
Schneider, P., Ehlers, J., Falco, E.~E., {et~al.} 1992, Gravitational Lenses,
  157

\bibitem[{Schneider {et~al.}(1998)Schneider, Van~Waerbeke, Jain, \&
  Kruse}]{schneider1998new}
Schneider, P., Van~Waerbeke, L., Jain, B., \& Kruse, G. 1998, Monthly Notices
  of the Royal Astronomical Society, 296, 873

\bibitem[{Schneider {et~al.}(2002)Schneider, van Waerbeke, Kilbinger, \&
  Mellier}]{schneider2002analysis}
Schneider, P., van Waerbeke, L., Kilbinger, M., \& Mellier, Y. 2002, Astronomy
  \& Astrophysics, 396, 1

\bibitem[{Schrabback {et~al.}(2015)Schrabback, Hilbert, Hoekstra, Simon, van
  Uitert, Erben, Heymans, Hildebrandt, Kitching, Mellier,
  {et~al.}}]{schrabback2015cfhtlens}
Schrabback, T., Hilbert, S., Hoekstra, H., {et~al.} 2015, Monthly Notices of
  the Royal Astronomical Society, 454, 1432

\bibitem[{Schrabback {et~al.}(2018)Schrabback, Schirmer, van Der~Burg,
  Hoekstra, Buddendiek, Applegate, Brada{\v{c}}, Eifler, Erben, Gladders,
  {et~al.}}]{schrabback2018precise}
Schrabback, T., Schirmer, M., van Der~Burg, R.~F., {et~al.} 2018, Astronomy \&
  Astrophysics, 610, A85

\bibitem[{Seppi {et~al.}(2024)Seppi, Comparat, Ghirardini, Garrel, Artis,
  S{\'a}nchez, Liu, Clerc, Bulbul, Grandis, {et~al.}}]{seppi2024srg}
Seppi, R., Comparat, J., Ghirardini, V., {et~al.} 2024, Astronomy \&
  Astrophysics, 686, A196

\bibitem[{Shan {et~al.}(2012)Shan, Kneib, Tao, Fan, Jauzac, Limousin, Massey,
  Rhodes, Thanjavur, \& McCracken}]{shan2012weak}
Shan, H., Kneib, J.-P., Tao, C., {et~al.} 2012, The Astrophysical Journal, 748,
  56

\bibitem[{Shan {et~al.}(2018)Shan, Liu, Hildebrandt, Pan, Martinet, Fan,
  Schneider, Asgari, Harnois-D{\'e}raps, Hoekstra, {et~al.}}]{shan2018kids}
Shan, H., Liu, X., Hildebrandt, H., {et~al.} 2018, Monthly Notices of the Royal
  Astronomical Society, 474, 1116

\bibitem[{{Smith} {et~al.}(2003){Smith}, {Peacock}, {Jenkins}, {White},
  {Frenk}, {Pearce}, {Thomas}, {Efstathiou}, \& {Couchman}}]{smith03}
{Smith}, R.~E., {Peacock}, J.~A., {Jenkins}, A., {et~al.} 2003, \mnras, 341,
  1311

\bibitem[{{Springel} {et~al.}(2001){Springel}, {White}, {Tormen}, \&
  {Kauffmann}}]{springel01b}
{Springel}, V., {White}, S.~D.~M., {Tormen}, G., \& {Kauffmann}, G. 2001,
  \mnras, 328, 726

\bibitem[{{Takahashi} {et~al.}(2012){Takahashi}, {Sato}, {Nishimichi},
  {Taruya}, \& {Oguri}}]{takahashi12}
{Takahashi}, R., {Sato}, M., {Nishimichi}, T., {Taruya}, A., \& {Oguri}, M.
  2012, \apj, 761, 152

\bibitem[{Tang \& Fan(2005)}]{tang2005effects}
Tang, J. \& Fan, Z. 2005, The Astrophysical Journal, 635, 60

\bibitem[{{Tessore} {et~al.}(2015){Tessore}, {Winther}, {Metcalf}, {Ferreira},
  \& {Giocoli}}]{tessore15}
{Tessore}, N., {Winther}, H.~A., {Metcalf}, R.~B., {Ferreira}, P.~G., \&
  {Giocoli}, C. 2015, \jcap, 10, 036

\bibitem[{To {et~al.}(2021)To, Krause, Rozo, Wu, Gruen, Wechsler, Eifler,
  Rykoff, Costanzi, Becker, {et~al.}}]{to2021dark}
To, C., Krause, E., Rozo, E., {et~al.} 2021, Physical review letters, 126,
  141301

\bibitem[{Toni {et~al.}(2024)Toni, Maturi, Finoguenov, Moscardini, \&
  Castignani}]{toni2024amico}
Toni, G., Maturi, M., Finoguenov, A., Moscardini, L., \& Castignani, G. 2024,
  Astronomy \& Astrophysics, 687, A56

\bibitem[{Umetsu(2020)}]{umetsu2020cluster}
Umetsu, K. 2020, The Astronomy and Astrophysics Review, 28, 1

\bibitem[{Veropalumbo {et~al.}(2014)Veropalumbo, Marulli, Moscardini, Moresco,
  \& Cimatti}]{veropalumbo2014improved}
Veropalumbo, A., Marulli, F., Moscardini, L., Moresco, M., \& Cimatti, A. 2014,
  Monthly Notices of the Royal Astronomical Society, 442, 3275

\bibitem[{Vikhlinin {et~al.}(2009)Vikhlinin, Kravtsov, Burenin, Ebeling,
  Forman, Hornstrup, Jones, Murray, Nagai, Quintana,
  {et~al.}}]{vikhlinin2009chandra}
Vikhlinin, A., Kravtsov, A., Burenin, R., {et~al.} 2009, The Astrophysical
  Journal, 692, 1060

\bibitem[{Werner {et~al.}(2023)Werner, Cypriano, Gonzalez, Mendes~de Oliveira,
  Araya-Araya, Doubrawa, Lopes~de Oliveira, Lopes, Vitorelli, Brambila,
  {et~al.}}]{werner2023s}
Werner, S., Cypriano, E., Gonzalez, A., {et~al.} 2023, Monthly Notices of the
  Royal Astronomical Society, 519, 2630

\bibitem[{White {et~al.}(2002)White, van Waerbeke, \&
  Mackey}]{white2002completeness}
White, M., van Waerbeke, L., \& Mackey, J. 2002, The Astrophysical Journal,
  575, 640

\bibitem[{Wittman {et~al.}(2006)Wittman, Dell’Antonio, Hughes, Margoniner,
  Tyson, Cohen, \& Norman}]{wittman2006first}
Wittman, D., Dell’Antonio, I., Hughes, J., {et~al.} 2006, The Astrophysical
  Journal, 643, 128

\end{thebibliography}
\end{document}